\documentclass[useAMS,referee,usenatbib]{biom}
\usepackage{graphicx}
\usepackage{url}

\def\bSig\mathbf{\Sigma}

\title[Fitting Double Hierarchical Models with INLA]{Fitting Double Hierarchical Models with the Integrated Nested Laplace Approximation}

\author
{Mabel Morales-Otero$^1$,
Virgilio G\'omez-Rubio$^{2,*}$\email{Virgilio.Gomez@uclm.es}, and
Vicente N\'u\~{n}ez-Ant\'on$^1$\\
$^1$Departamento de M\'etodos Cuantitativos,\\Universidad del Pa\'is Vasco UPV/EHU, Bilbao, Spain\\
$^2$Departamento de Matem\'aticas, Escuela T\'ecnica Superior de Ingenieros Industriales,\\ Universidad de Castilla-La Mancha (UCLM), Albacete, Spain
}

\begin{document}







\doi{}


\label{firstpage}


\begin{abstract}


Double hierarchical generalized linear models (DHGLM) are a family
of models that are flexible enough as to model hierarchically the 
mean and scale parameters. In a Bayesian framework,
fitting highly parameterized hierarchical models is challenging
when this problem is addressed using typical Markov chain Monte Carlo (MCMC)
methods due to the potential high correlation between different parameters
and effects in the model. The integrated nested Laplace approximation (INLA)
could be considered instead to avoid dealing with these problems. However,
DHGLM do not fit within the latent Gaussian Markov random field (GMRF) models
that INLA can fit.

In this paper we show how to fit DHGLM with INLA by combining INLA and
importance sampling (IS) algorithms. In particular, we will illustrate how to split
DHGLM into submodels that can be fitted with INLA so that the remainder of the
parameters are fit using adaptive multiple IS (AMIS) with the aid of the
graphical representation of the hierarchical model.  This is illustrated using
a simulation study on three different types of models and two real data examples.

\end{abstract}

%



\maketitle


%

\section{Introduction}
\label{s:intro}

Double hierarchical generalized linear models \citep[DHGLM,][]{LeeNelder:2006}
provide a unique approach to modelling highly structured datasets allowing for
additional flexibility, particularly when modelling the dispersion parameters.
The class of DHGLM encompasses a large variety of models such as standard
generalized linear models (GLM), mixed effects models, random coefficient
models, semi-parametric models and many others.  A typical DHGLM includes a
linear mixed-effects term to model the mean as well as several terms to model
the scale parameters of the likelihood and/or random effects present in the
model.

Estimation of DHGLM can be approached in different ways. \cite{LeeNelder:2006}
propose the use of the H-likelihood for model fitting and \cite{ronnegard2010}
use penalized quasi-likelihood. Bayesian inference on DHGLM allows us to
estimate the different effects and parameters in the model and estimate their
uncertainty by means of the joint posterior distribution.

Because of the different structured terms and effects in a DHGLM, model fitting
can become a daunting task. Popular methods such as Markov chain Monte Carlo
(MCMC) could take a long time and many simulations to converge. The integrated
nested Laplace approximation \citep[INLA, ][]{Rueetal:2009} approach is 
an appealing option because of its short computation time. 
However, DHGLM are a class of models that are not
currently implemented in the \texttt{R-INLA} package for the \texttt{R}
programming language.

In order to fit models that cannot be currently implemented in INLA,
given their specific structure, there have been some developments to combine INLA
with other methods such as Markov chain Monte Carlo (MCMC) methods
\citep{GomezRubioRue:2018,GomezRubio:2020}. \cite{Berildetal:2021} replace MCMC methods by
importance sampling (IS) and adaptive multiple IS (AMIS) to achieve model fitting in
a faction of INLA within MCMC.

In this work, we propose fitting DHGLM with the use of the AMIS-INLA approach
\citep{Berildetal:2021}. We show how this class of models can be fitted in this
way, providing specific details for the implementation of the algorithms in the
cases where variables following Gaussian, Poisson and negative binomial
distributions are modelled. \cite{LeeNoh:2012} describe ways of modeling 
the variance of the random effects for DHGLM. We have focused on modeling 
the precision instead, but the approach presented here can also be 
used to model the variance or standard deviation when needed.

This paper is structured as follows. Section~\ref{s:dhglm} describes DHGLM.
Section~\ref{s:inla} summarizes the integrated nested Laplace approximation for Bayesian 
inference. Section ~\ref{s:modelfitting} describes model fitting combining AMIS and INLA, and
Section~\ref{s:simulation} includes a simulation study to assess the behavior of our proposed
methodology. Section~\ref{s:examples} illustrates the usefulness of the proposed method by applying them 
to two real data examples. Finally, Section~\ref{s:discuss} provides a discussion and summary of our
results.

\section{Double Hierarchical Generalized Linear Models}
\label{s:dhglm}

Suppose $Y_{i}$, for $i=1, \dots, n$ are random variables following a distribution from the exponential family \citep{mccullagh:1989}. That is, their probability distribution function can be written as:

\begin{eqnarray*}
 f(y_{i}; \theta_{i}, \phi_{i}) = \rmn{exp} \left\{ \frac{y_{i}\theta_{i}-b(\theta_{i})}{\theta_{i}} + c(y_{i},\phi_{i}) \right\},
\end{eqnarray*}
where $y_{i}$ is the observation corresponding to the variable $Y_{i}$, $\theta_{i}$ is a vector of parameters, $\phi_{i}$ is a known positive constant value labelled as the scale or dispersion parameter, and $b(.)$ and $c(.)$ are given known functions.

It is known that ${\rmn E}[Y_{i}]= \mu_{i} = b^{\prime}(\theta_{i})$ and that $\rmn{Var}[Y_{i}]=\theta_{i}V(\mu_{i})$, with $V(\mu_{i})=b^{\prime\prime}(\theta_{i})$ being a variance function. Different forms for $\phi_{i}$ and $V(\mu_{i})$ for some known distributions are included in Table~\ref{tab:exp_family}, where $\sigma^{2}$ is the variance parameter for the normal distribution, $n_{i}$ is the number of observations on each trial of the binomial distribution and $k$ the dispersion  parameter or size of the negative binomial distribution.

\begin{table}[h!]
\caption{Different form of $\phi_{i}$ and $V(\mu_{i})$  for some known distributions.}
\label{tab:exp_family}
\centering
\begin{tabular}{c|c|c}
\hline
Distribution & $\phi_{i}$ &  $V(\mu_{i})$ \\
\hline
Normal            & $\phi_{i}=\sigma^2$ & $V(\mu_{i})=1$  \\
Poisson           & $\phi_{i}=1$        & $V(\mu_{i})= \mu_{i}$  \\
Negative binomial & $\phi_{i}=1$        & $V(\mu_{i})= \mu_{i} + k^{-1}\mu^2_{i}$  \\
Binomial          & $\phi_{i}=1$        & $V(\mu_{i})= \mu_{i}\left(\frac{n_{i}-\mu_{i}}{n_{i}}\right)$  \\
\end{tabular}
\end{table}

For example, for a variable $Y_{i}$ having a negative binomial distribution, its probability mass function can be specified as:

\begin{eqnarray*}
 f(y_{i}; p_{i}, k) = P(Y_{i}=y_{i}) = {{y_{i} + k + 1} \choose {y_{i}}} p_{i}^{k}(1-p_{i})^{y_{i}},
\end{eqnarray*}
where $p_{i}$ is the probability of success on a Bernoulli trial and $y_{i}$ would represent the number of failures before the $k$-th success occurs.
The mean is ${\rmn E}[Y_i]= \mu_i= k \left(\frac{1-p_{i}}{p_{i}}\right)$ and the variance is ${\rmn Var}[Y_i]=k \left(\frac{1-p_{i}}{p_{i}^{2}}\right)=\mu_i+k^{-1}\mu_i^{2}$.
If the parameter $k$ is considered fixed, this distribution belongs to the exponential family \citep[see][]{agresti:2002}.

A generalized linear model (GLM) \citep{mccullagh:1989} is defined when a regression model is specified for the mean via a link function $g(.)$, obtaining a linear predictor for the $i$-th observation, so that:
\begin{equation}
g(\mu_{i}) = \eta_i = \bld{X}_{i}^{\top}\bm{\beta},
\label{eq:glm}
\end{equation}
where $\bld{X}_{i}$ is a vector of explanatory variables and $\bm{\beta}$ is a vector of unknown regression parameters to be estimated.

GLM were further extended by \cite{LeeNelder:2006} by proposing the DHGLM,
which are specified given a set of two random effects
$(\bld{u}^{(\bm{\mu})},\bld{u}^{(\bm{\phi})})$, so that the conditional mean and variance
of the response variables $Y_i$ are
${\rmn E}[Y_{i}|\bld{u}^{(\bm{\mu})},\bld{u}^{(\bm{\phi})}]=\mu_i$ and
$\rmn{Var}[Y_{i}|\bld{u}^{(\bm{\mu})},\bld{u}^{(\bm{\phi})}]=\phi_{i}V(\mu_i)$,
respectively, for $i=1, \dots, n$. The
random effects depend on the variance (or precision) parameters $\bm{\lambda}$ and
$\bm{\alpha}$, i.e., $(\bld{u}^{(\bm{\mu})}(\bm{\lambda}),\bld{u}^{(\bm{\phi})}(\bm{\alpha}))$.  Here,
regression models for the mean, for the dispersion parameters and for the
parameters of the random effects are specified, so that:

\begin{equation}
\begin{array}{rcl}
    g^{(\bm{\mu})}(\mu_{i}) & = & \bld{X}_{i}^{\top(\bm{\mu})}\bm{\beta}^{(\bm{\mu})} + \bld{Z}_{i}^{\top(\bm{\mu})}u_{i}^{(\bm{\mu})}\\
    g^{(\bm{\lambda})}(\lambda_{i}) & = & \bld{X}_{i}^{\top(\bm{\lambda})}\bm{\beta}^{(\bm{\lambda})} \\
    g^{(\bm{\phi})}(\phi_{i}) & = & \bld{X}_{i}^{\top(\bm{\phi})}\bm{\beta}^{(\bm{\phi})} + \bld{Z}_{i}^{\top(\bm{\phi})}u_{i}^{(\bm{\phi})}\\
    g^{(\bm{\alpha})}(\alpha_{i}) & = & \bld{X}_{i}^{\top(\bm{\alpha})}\bm{\beta}^{(\bm{\alpha})}, \,
\end{array}
\label{eq:dhglm}
\end{equation}
where $\bld{X}_{i}^{\top(.)}$ is the $i$-th row of the design matrix $\bld{X}^{(.)}$ for $\bm\mu, \bm\lambda, \bm\phi$ and $\bm\alpha$, $\bld{Z}_{i}^{\top(.)}$ is the $i$-th row of the design matrix $\bld{Z}^{(.)}$ for $\bm\mu, \bm\phi$, and $\bm{\beta}^{(.)}$ is a vector of unknown coefficients to be estimated for $\bm\mu, \bm\lambda, \bm\phi$ and $\bm\alpha$, respectively.

As we have previously mentioned, estimation of this model can be done by using
the H-likelihood proposed by \cite{LeeNelder:2006} and also penalized
quasi-likelihood proposed by \cite{ronnegard2010}. Bayesian methods have been
widely employed to fit highly parameterized hierarchical models in the context
of DHGLM \citep[see, for example,][and the references therein]{dalmatian}. In
\cite{cepeda:2018} and \cite{morales:2021}, the authors use MCMC methods to
fit generalized overdispersion models, where regression structures depending
on some covariates are specified both for the mean and for the dispersion
parameters.

DHGLM aim at modeling the dispersion parameter $\bm\phi$ using a linear term
with fixed and random effects, perhaps after conveniently using a link function
$g(\cdot)$, so that $g(\phi)= \eta^{'}_i$, with $\eta^{'}_i$ being a linear
predictor.  These ideas can be extended to the case of dispersion parameters of
Gaussian distributed random effects as well.

Although most authors have developed a model on the log-variance
$\log(\sigma^2)$, in our case, we prefer to define the model on the
log-precision, $\log(\tau)$, with $\sigma^2 = 1 / \tau$, as the parameterization
of many distributions is in terms of the precision. In addition, given that
$\log(\sigma^2) = -\log(\tau)$, in terms of model fitting, the only difference
between the two approaches is a sign shift in the coefficients of the fixed
effects and the random effects.

\section{Integrated Nested Laplace Approximation}
\label{s:inla}

The integrated nested Laplace approximation (INLA) was first proposed by
\cite{Rueetal:2009} to provide fast approximate Bayesian inference for latent
Gaussian Markov random field (GMRF) models.  Given a set of $n$ observed
variables $\mathbf{Y} = (Y_1, \ldots, Y_n)$, usually with a distribution from the
exponential family, the density of $Y_i,\ i=1,\ldots,n$ may depend on
some hyperparameters $\bm{\theta_1}$. In addition, the mean of $Y_i$, ${\rmn E}[Y_i]$, will
be linked to a linear predictor $\eta_i$ on the covariates using a convenient
link function $g(\cdot)$ so that $g[{\rmn E}(Y_i)] = \eta_i$.

The linear predictor may include different terms, as fixed and/or random effects, so that the distribution of all these terms is a GMRF with zero mean and
precision matrix $Q(\bm{\theta_2})$, that may depend on some other hyperparameters
$\bm{\theta_2}$. To simplify notation, we will often use $\bm{\theta} = (\bm{\theta_1}, \bm{\theta_2})$ to refer to the vector of hyperparameters.
In addition, the vector of latent effects will be denoted by $\bm{\kappa}$.

In a Bayesian framework, the aim is to compute the posterior distribution of
the latent effects and hyperparameters, $\pi(\bm\kappa, \bm\theta \mid
\mathcal{D})$, to make inference about them. Here, $\mathcal{D}$ represents the
available data, which will include the observed response $\mathbf{y} = (y_1,\ldots,y_n)$
and, possibly, other covariates required to define the fixed and random effects
in the latent GMRF. Using Bayes' rule, this joint posterior distribution can be
written as:

$$
\pi(\bm\kappa, \bm\theta \mid \mathcal{D}) \propto
L(\bm\kappa, \bm\theta \mid \mathcal{D}) \pi(\bm\kappa, \bm\theta)
$$
\noindent
Here, $L(\bm\kappa, \bm\theta \mid \mathcal{D})$ represents the likelihood
of the data, while $\pi(\bm\kappa, \bm\theta)$ is the joint prior distribution
of the latent effects and hyperparameters. This is often expressed as
$\pi(\bm\kappa, \bm\theta) = \pi(\bm\kappa \mid \bm\theta) \pi(\bm\theta)$.
Note that $\pi(\bm\kappa \mid \bm\theta) $ is a GMRF and $\pi(\bm\theta)$ is
often defined as the product of univariate distributions as hyperparameters
are considered to be independent a priori.

The joint posterior distribution $\pi(\bm\kappa, \bm\theta \mid \mathcal{D})$ is
often highly multivariate and difficult to estimate. For this reason,
\cite{Rueetal:2009} focus on estimating the marginal posterior distributions of
the latent effects and hyperparameters. In this way, approximations
$\tilde{\pi}(\kappa_j \mid \mathcal{D})$  and $\tilde{\pi}(\theta_l \mid
\mathcal{D})$ to $\pi(\kappa_j \mid \mathcal{D})$  and $\pi(\theta_l \mid
\mathcal{D})$, respectively, are obtained.

In addition, INLA can be used to obtain an approximation to the marginal likelihood
of the model, $\pi(\mathcal{D})$, which is often difficult to compute.
Other important quantities for model selection and model choice
are available in the R-INLA package that implements the INLA method.

As discussed in the next section, INLA cannot fit DHGLM directly, but
INLA can be embedded into the model fitting process to be able to easily fit these models.

\section{Model Fitting}
\label{s:modelfitting}

DHGLM do not fall into the class of models that INLA can fit due to their
particular structure that includes different hierarchies on the mean and
scale parameters. However, as explained below, DHGLM can be expressed as conditional
latent GMRF models after conditioning on some model parameters.  This idea of
fitting conditional models with INLA has been exploited by several authors
\citep[see, for example,][]{GomezRubioRue:2018} to increase the number of
models that can be fit with INLA.

In particular, the vector of hyperameters $\bm\theta$ can be decomposed
into two sets of parameters $\bm\theta_c$ and $\bm\theta_{-c}$, so that
the model, conditional on $\bm\theta_c$, can be fit with INLA. The posterior
distribution of $\bm\theta_c$ can then be expressed as

$$
\pi(\bm\theta_c \mid \mathcal{D}) \propto \pi(\mathcal{D} \mid \bm\theta_c)
\pi(\bm\theta_c)
$$

\noindent Here, $\pi(\bm\theta_c)$ is the prior on $\bm\theta_c$, which is known,
and $\pi(\mathcal{D} \mid \bm\theta_c)$ is the \textit{conditional} (on $\bm\theta_c$) marginal likelihood,
as this is obtained after integrating out all the other hyperparameters and latent effects. 
This quantity can be easily obtained with INLA, so that the posterior distribution of $\pi(\bm\theta_c \mid \mathcal{D})$
can be estimated.

Regarding the other hyperparameters $\bm\theta_{-c}$ and the latent
effects, their marginal posterior distributions can be obtained by noting that

$$
\pi(\cdot \mid \mathcal{D}) = \int \pi(\cdot, \bm\theta_c \mid \mathcal{D}) d\bm\theta_c = \int \pi(\cdot \mid \bm\theta_c,  \mathcal{D}) \pi(\bm\theta_c \mid \mathcal{D}) d\bm\theta_c
$$
\noindent
The conditional posterior marginal $\pi(\cdot \mid \bm\theta_c,  \mathcal{D})$
is provided by INLA when fitting the model, conditional on $\bm\theta_c$.

In practice, an approximation to $\pi(\cdot \mid \mathcal{D})$ is obtained
by weighing the posterior conditional marginals. That is,

$$
\tilde{\pi} (\cdot \mid \mathcal{D}) =
\sum_{m=1}^{M} \tilde\pi(\cdot \mid \bm\theta_c^{(m)},  \mathcal{D}) w_m
$$
\noindent
Here, $M$ represents a number of ensemble of values of $\bm\theta_c^{(m)}$,
$\{\bm\theta_c^{(m)}\}_{m=1}^M$, that is used for numerical integration.
In addition, $w_m$ are weights that can be computed in different ways, depending
on how the values of $\bm\theta_c$ have been obtained.

In this regard, \cite{GomezRubioRue:2018} use the Metropolis-Hastings
algorithm to estimate the distribution of $\bm\theta_c$, and also use the resulting
values to estimate the remainder of the latent effects and hyperparameters. This
algorithm requires fitting a model with INLA at each iteration of the Metropolis-Hastings algorithm, which makes it less appealing in practice.

Similarly, \cite{Berildetal:2021} use the importance sampling (IS) algorithm
instead, which can be run in parallel and provides reduced computing
times. In this particular case, samples of $\bm\theta_c$ are obtained using an
importance distribution $s(\cdot)$ to obtain $\{\bm\theta_c^{(m)}\}_{m=1}^M$.
For each value $\bm\theta_c^{(m)}$, a conditional model is fit with INLA
so that integration weights $w_m$ are obtained as follows:

$$
w_m \propto \frac{\pi(\mathcal{D} \mid \bm\theta_c) \pi(\bm\theta_c^{(m)})}{s(\bm\theta_c^{(m)})}
$$

Weights are re-scaled so that they sum up to one. Furthermore,
\cite{Berildetal:2021} describe the use of the adaptive multiple importance
sampling \citep[AMIS,][]{Corneutetal:2012} algorithm that provides a more robust
sampling method that updates the importance distribution $s(\cdot)$.

Regarding model fitting of DHGLM with INLA, we will use IS and AMIS with INLA
by conditioning on some of the model hyperparameters or latent effects.  These
will depend on the way in which the DHGLM is defined.  Both
\cite{GomezRubioRue:2018} and \cite{Berildetal:2021} discuss different
approaches on how to best select the parameters in $\bm\theta_c$. In the
simplest cases, the choice of $\bm\theta_c$ will be clear as just a few
parameters will need to be fixed to obtain a conditional latent GMRF model. For
highly structured models, it may happen that after conditioning on some
hyperparameters or latent effects, two or more conditionally independent submodels appear
\citep[see, for example,][]{Lazaroetal:2020}. These submodels can be fit
independently with INLA. All the different cases are illustrated in
Section~\ref{s:simulation}, where different simulations studies are developed
in detail on different types of models.

However, in order to provide a more general approach to the choice of
$\bm\theta_c$, we propose the use of a graphical representation of the model.
This graphical model encodes conditional independence relationships among the
model parameters, so that its structure can be exploited to be able to select
the best possible choice of the parameters to be included in $\bm\theta_c$ \citep[see, for
example,][]{Cowelletal:1999}. See Section~\ref{s:summary} for more details
and a thorough discussion about this graphical representation using the examples 
in the simulation study conducted in Section~\ref{s:simulation}.

Regarding the sampling distribution for $\bm\theta_c$, \cite{Berildetal:2021}
suggest choosing a multivariate Gaussian distribution or a multivariate $t$
distribution with a small number of degrees of freedom for continuous
variables.  Note that some of the variables in $\bm\theta_c$ may need to be
re-scaled (e.g., a precision will be sampled in the log scale). Hence, the mean
and precision of these distributions are updated at each adaptive step.  For
discrete variables, the choice is not so clear. When the variables are
dichotomous, \cite{Berildetal:2021} suggest using a binomial distribution for
each of them, so that the probabilities depend on some fixed effects (which are
the parameters updated after each adaptive step).

The choice of the parameters of the sampling distribution is crucial to
obtain a good performance of the proposed methodology. The initial parameters of the distribution
could be based on sample statistics as rough estimates. For example,
for continuous data, when the sampling distribution is a multivariate normal,
the mean can be obtained by obtaining the sample mean of the parameters
and the precision can be diagonal with \textit{large} values in the diagonal.
Here \textit{large} must be put into context according to the scale of the
parameters. Too large values of the precision will imply that the parameter
space is not conveniently explored, while too small values will imply that
samples with a very small posterior density will be sampled too often. In both
cases, bad estimates will be obtained at the adaptive steps that can result in
the algorithm requiring more steps to produce reliable estimates. This issue
is thoroughly discussed in the simulation study in Section~\ref{s:simulation}
and the examples in Section~\ref{s:examples}.

In addition, as a general guidance, the conditional model can be fit with INLA given
the set of possible values for the mean of the sampling distribution before
running AMIS with INLA. Different sets of values can be tested and the marginal
likelihoods compared. The one with the highest value of the marginal likelihood
may be a better candidate as it improves model fitting. This will help to be able to
choose an initial sampling distribution whose mode is close to the posterior
mode of $\bm\theta_c$, so that less adaptive steps (and, hence, simulations)
are required to obtain good estimates.

Note that, because these values will help to define the sampling distribution,
they can be based on summary statistics from the data or maximum likelihood
estimates of conveniently chosen models.

Another way of assessing the performance of IS with INLA is to compute the
effective sample size and conduct graphical diagnostics, as discussed in \cite{Berildetal:2021}.
The effective sample size can be estimated as

$$
n_e = \frac{(\sum_{m=1}^M w_m)^2}{\sum_{m=1}^M w^2_m} .
$$
\noindent
Note that this effective sample size will be the same for all the components
of $\bm\theta_c$ as it is only based on the weights and not on the sampled
values.

Graphical diagnostics can be produced for each variable in $\bm\theta_c$ by
re-ordering the sampled values in ascending order and comparing the estimated
cumulative  probability (i.e., the cumulative sum of the re-ordered weights)
with the empirical cumulative probabilities $1/M, \ldots, M/M$, respectively A
straight line means that the estimated posterior marginal of that
parameter is reliable.

Monitoring the convergence of the algorithm could be conducted in a number of
ways. First of all, the effective sample size could be computed and the
algorithm can be stopped once the desired sample size has been achieved. The
conditional marginal likelihood fitted at the mean of the sampling distribution
after each adaptive step could also be monitored to assess whether it keeps
increasing or approaches a certain value (at this point the algorithm
can be stopped). It is worth noting that more samples could be obtained when 
needed by simply resuming the simulations using
updated estimates of the parameters of the sampling distribution.

\section{Simulation study}
\label{s:simulation}

In this section we develop three different simulation studies to illustrate
model fitting of hierarchical models with different structures. In
Section~\ref{s:poisson} we fit a Poisson log-linear model with random effects, in
which the log-precision of the random effects is modeled using a linear term; in
Section~\ref{s:negbin} we fit a negative binomial model in which the
log-size parameter is modeled using a linear term; and in Section~\ref{s:gaussian}
we fit a Gaussian model to grouped data in which the log-precision of each
group is modeled using a linear mixed-effects model. In all cases, models are
fit using MCMC and AMIS with INLA. IS with INLA has not been considered because
\cite{Berildetal:2021} show that, in general, AMIS-INLA has a better performance than IS
with INLA.

The aim of these simulation studies is twofold. On the one hand, we would like
to illustrate the way in which IS and AMIS with INLA can be implemented and how
the conditioning effects $\bm\theta_c$ can be chosen. On the other hand, it is important
to compare the results obtained with these methods to a gold standard.
In our case, we have fitted the models using Markov chain Monte Carlo
\citep{Brooksetal:2011} using the JAGS software via the \texttt{R}-package \texttt{rjags} \citep{rjags}.

\begin{figure}
\centering
\includegraphics[scale=0.9]{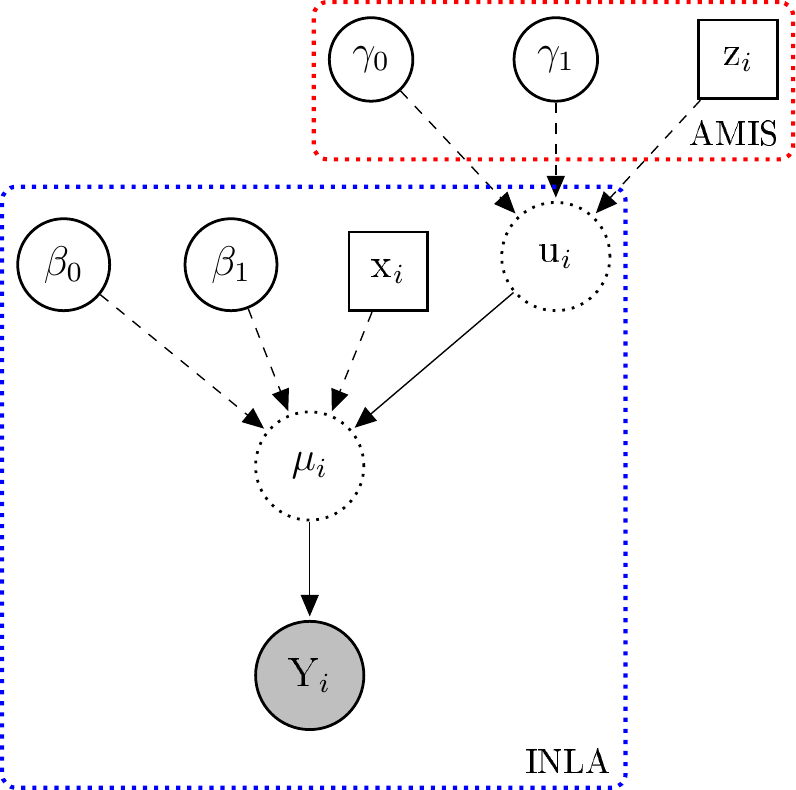} \ \ \ \ \ \
\includegraphics[scale=0.9]{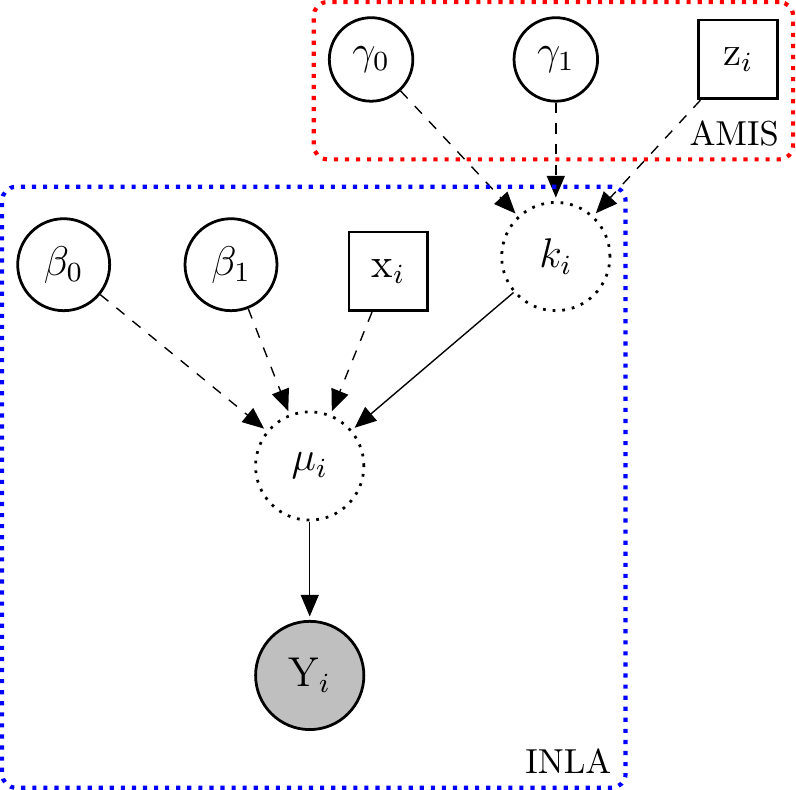}

\

\

\includegraphics[scale=1]{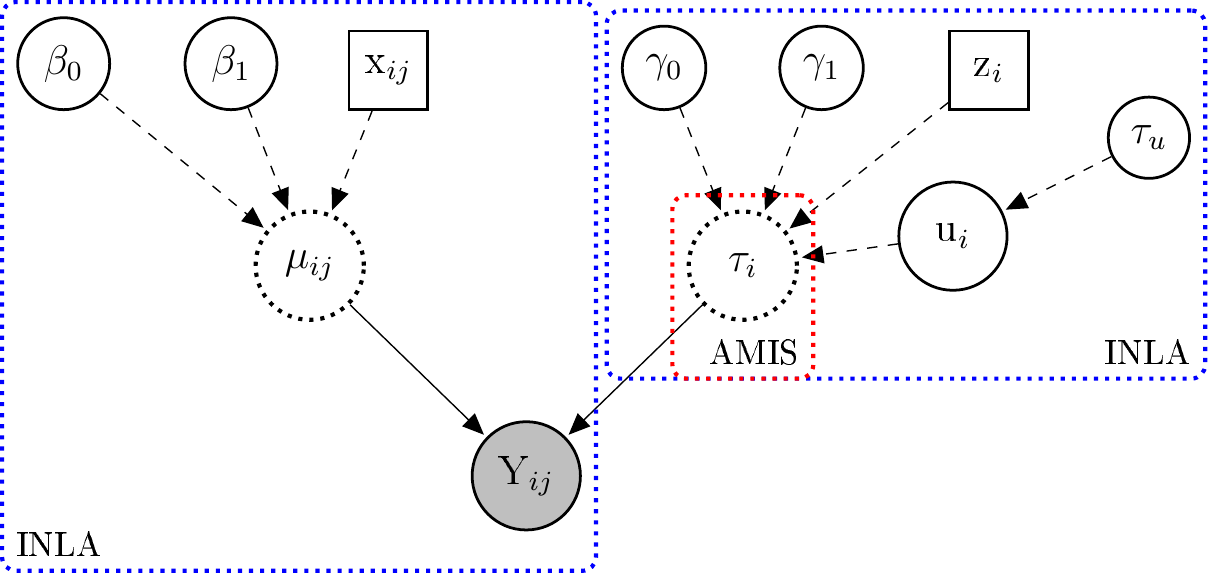}
\caption{Graphical representation of the models fit in the simulation study in Section~\ref{s:simulation}: Poisson model with random effects (top-left), negative binomial model with regression on the log-sizes (top-right) and Gaussian model with regression model on the likelihood log-precisions (bottom). Nodes in a shaded solid circle represent the observed data, nodes in a white solid circle represent model effects and parameters and nodes in a dotted white circle represent
deterministic nodes (i.e., their values are fully determined by the values at
their parent nodes). Parameters fit with AMIS with INLA are in a red dotted box, whereas the conditional model fit with INLA is in a blue dotted box.}
\label{fig:graphmodels}
\end{figure}

Figure~\ref{fig:graphmodels} shows the representation of these models as
graphical models. In addition to the different elements of the model, the
conditioning parameters have been highlighted (using a red dotted box)
to illustrate which parameters are  estimated using AMIS. The marginals
of all the other parameters are obtained by averaging the conditional marginals
resulting after fitting the conditional models with INLA. Nodes in a shaded solid
circle represent the observed data, nodes in a white solid circle represent
model effects and parameters and nodes in a dotted white circle represent
deterministic nodes (i.e., their values are fully determined by the values at
their parent nodes).


When implementing AMIS, the importance distribution $s(\cdot)$ is assumed as a multivariate
Gaussian in all examples. Note that this means that some parameters may be
transformed so that simulations are feasible. For example, precisions will be
sampled in the log-scale, so that samples from the log-precision are obtained.
In all cases we have computed results using a very vague distribution (with zero
mean and large precision) and another distribution based on a rough estimate of
the parameters of the sampling distribution from the observed data.
This should reduce the number of iterations required to obtain a
reliable model fitting. In all cases, the estimates from AMIS with INLA are
compared with MCMC estimates.

In all the examples presented below the same number of simulations has been
used. When fitting the model using AMIS with INLA, 5000 iterations have been
used in the initial step, followed by 10 new adaptive steps with 1000
simulations each. For MCMC, a burn-in of 10000 simulations is used, plus 100000
simulations of which only one in 100 is retained, leading to a final number of
1000 samples.  In addition, in the Gaussian example in
Section~\ref{s:gaussian}, different scenarios have been tested (see below for
details).  Finally, simulations have been carried out on a Linux Ubuntu 18.2
cluster using 60 cores Intel(R) Xeon(R) CPU E5-2683 v4 @ 2.10GHz.

\subsection{Poisson model with random effects with different precisions}
\label{s:poisson}

The first simulation study is based on a Poisson log-linear model
with fixed and random effects, so that the precision of the random effects
is modeled using a linear term with covariates. In particular,
the model is

$$
\begin{array}{rcl}
Y_i &\sim & \rmn{Poi}(\mu_i),\ i=1,\ldots, n\\
\log(\mu_i) &= & \beta_0 + \beta_1 x_i + u_i\\
u_i &\sim& N(0, \tau_i)\\
\log(\tau_i) &=& \gamma_0 + \gamma_1 z_i\\
\beta_0,\beta_1 &\sim& N(0, 0.001)\\
\gamma_0,\gamma_1 &\sim& N(0, 0.001)\\
\end{array}
$$
\noindent
Note that the Gaussian distribution $N(\cdot,\cdot)$ is defined in terms
of the mean and precision so that $\tau_i$ represents the precision of the Gaussian distribution of the random effects. A Poisson distribution
with random effects is often used to model overdispersed data \citep{QuinteroSarmientoetal:2012}.

This model can be expressed as a latent GMRF by conditioning on $\bm\theta_c = \bm\gamma
= (\gamma_0, \gamma_1)$, resulting in a Poisson model with random effects
with different precisions. This is illustrated in the graphical representation
of the model in  Figure~\ref{fig:graphmodels} (top-left plot). Hence, this model will be fitted using AMIS with INLA and
values of $\bm\gamma$ will be obtained by simulation. Estimates of the
posterior distribution can be obtained by using importance weights
and the posterior marginals of $\beta_0$ and $\beta_1$ will be obtained by
weighting their conditional marginals.

For the simulated data, we have used $n=1000$, $\beta_0=1$, $\beta_1 = 0.25$,
$\gamma_0 = 0$ and $\gamma_1 = 0.5$. Covariate $x_i$ has been simulated using a
uniform distribution between 0 and 1, and covariate $z_i$ has been simulated
using a standard Gaussian distribution. Once these values have been set,
the observed value $y_i$ has been obtained by sampling from a Poisson distribution
with the resulting mean $\mu_i$.

The sampling distribution for $\bm\gamma$ is a bivariate Gaussian distribution.
The initial value of the mean is vector $(0, 0)$ and the initial value of the
variance matrix is a diagonal matrix with entries equal to 5 in the diagonal.
This choice provides ample initial variability to explore the parametric space
of $\bm\gamma$ conveniently so that accurate estimates are obtained at the
adaptive and final steps of AMIS with INLA.

\begin{figure}[h!]
\caption{Posterior marginals of the estimated parameters obtained by fitting the Poisson model with random effects, using both the MCMC and AMIS-INLA methods. Vertical lines represent the actual values of the parameters used when simulating the data.}
\includegraphics[width=\linewidth]{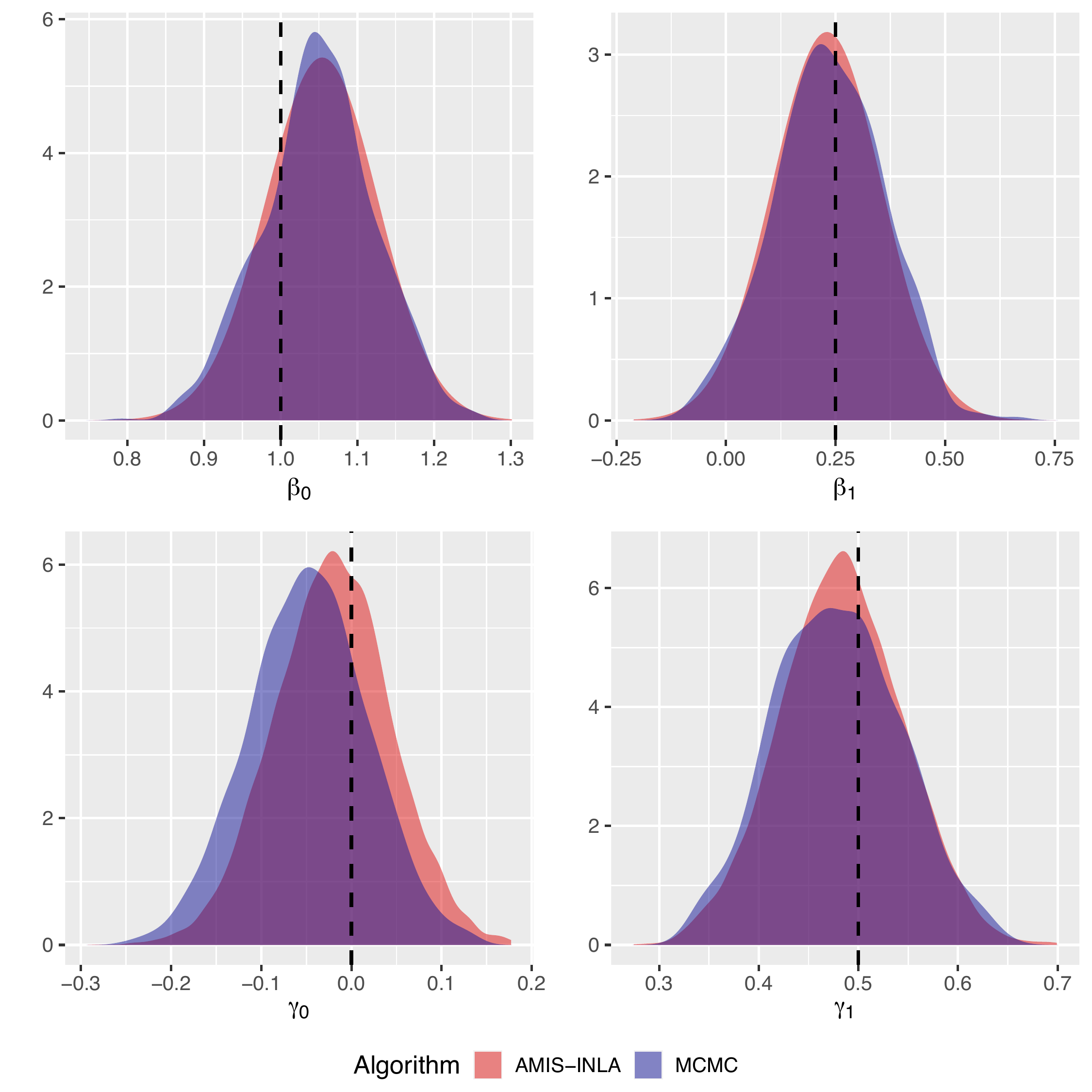}
\label{fig:poisson}
\end{figure}

Table~\ref{tab:poisson} summarizes the estimates using the different methods
and Figure~\ref{fig:poisson} shows the posterior marginal estimates obtained
with both methods. Here, the dashed vertical lines represent the true values of
the parameters specified for the simulated data. As can be seen, the
estimates obtained with AMIS with INLA and MCMC are very similar.  The
effective sample size $n_e$ obtained with AMIS with INLA in this case is
9900.914.

\begin{table}[h!]
\caption{Summary of the estimates of the Poisson model with random effects with different precisions used in the simulation study.}
\label{tab:poisson}
\centering
\begin{tabular}{c|c|cc|cc}
 & & \multicolumn{2}{|c|}{AMIS} & \multicolumn{2}{|c}{MCMC}\\
\hline
Parameter & True value & Mean & St. dev. & Mean & St. dev.\\
\hline
$\beta_0$ & 1 & 1.0531 & 0.0736 & 1.049 & 0.0729 \\
$\beta_1$ & 0.25 & 0.2302 & 0.1254 & 0.2347 & 0.1253 \\
$\gamma_0$ & 0 &  -0.0210 & 0.0655 & -0.0484 & 0.0654 \\
$\gamma_1$ & 0.5 & 0.4830 & 0.0622 & 0.4787 & 0.0636 \\
\end{tabular}
\end{table}

\subsection{Negative binomial with different sizes}
\label{s:negbin}

The negative binomial distribution is also used to model overdispersed
count data. In this simulation study the logarithm of the size parameter $k_i$ of the negative binomial
distribution depends on a linear term with covariates, which in turns
makes the probability to be different accross observations. In particular,
the model is as follows:

$$
\begin{array}{rcl}
Y_i &\sim & \rmn{NB}(p_i, k_i)\\
p_i & = & \frac{k_i}{k_i + \mu_i} \\
\log(\mu_i) &= & \beta_0 + \beta_1 x_i\\
\log(k_i) &=& \gamma_0 + \gamma_1 z_i\\
\beta_0,\beta_1 &\sim& N(0, 0.001)\\
\gamma_0,\gamma_1 &\sim& N(0, 0.001)\\
\end{array}
$$

Similarly, as in the previous example, this model can be expressed as a latent
GMRF by conditioning on $\bm\theta_c = \bm\gamma = (\gamma_0, \gamma_1)$,
resulting in a negative binomial model with different sizes.  This is
illustrated in the graphical representation of the model in
Figure~\ref{fig:graphmodels} (top-right plot). When fitting the model with AMIS
with INLA, values of $\bm\gamma$ will be obtained by simulation and their
estimates will be computed using the importance weights.  The posterior
marginals of $\beta_0$ and $\beta_1$ will be obtained by weighting their
conditional marginals.

For our study, $n=500$ observations have been simulated. Covariate $x_i$
is simulated from a uniform between 10 and 20, and covariate $z_i$ has been
simulated from a uniform between 0 and 20.  Values of $z_i$ have then been
standardized before simulating the data. Regarding the model parameters, we
have used $\beta_0=1$, $\beta_1 = 0.25$, $\gamma_0 = 0$ and $\gamma_1
= 5$. Once the mean and size of the negative binomial have been computed, the
values of the response variable have been sampled using a negative binomial
distribution.

Likewise, as in the previous simulation study, the sampling distribution for
$\bm\gamma$ is a bivariate Gaussian distribution.  The initial value of the
mean is vector $(0, 0)$ and the initial value of the variance matrix is a
diagonal matrix with entries equal to 5 in the diagonal. This a convenient
choice for this example as well and it provides good estimates of the
model parameters (see below).

\begin{figure}[h!]
\caption{Posterior marginals of the estimated parameters obtained by fitting the negative binomial model with different sizes, using both the MCMC and AMIS-INLA methods. Vertical lines represent the actual values of the parameters used when simulating the data.}
\includegraphics[width=\linewidth]{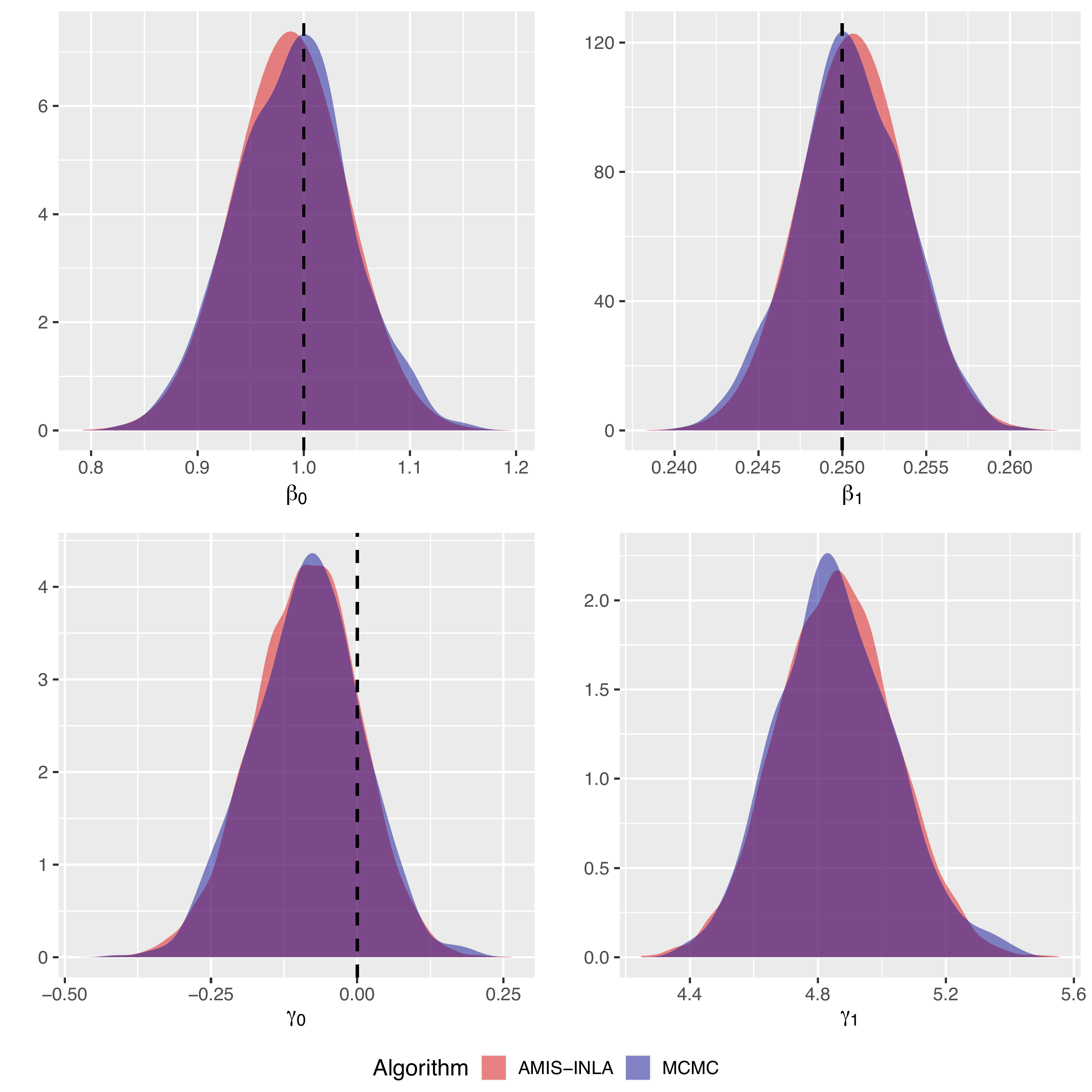}
\label{fig:negbin}
\end{figure}

Table~\ref{tab:negbin} summarizes the estimates using the different methods and
Figure~\ref{fig:negbin} shows the posterior marginal estimates obtained with
both methods.  As can be seen, the estimates obtained with AMIS with INLA
and MCMC are very similar.  The effective sample size $n_e$ obtained with AMIS
with INLA in this case is 9737.075.

\begin{table}[h!]
\caption{Summary of the estimates of the negative binomial model with different sizes used in the simulation study.}
\label{tab:negbin}
\centering
\begin{tabular}{c|c|cc|cc}
 & & \multicolumn{2}{|c|}{AMIS} & \multicolumn{2}{|c}{MCMC}\\
\hline
Parameter & True value & Mean & St. dev. & Mean & St. dev.\\
\hline
$\beta_0$ & 1 & 0.9875 & 0.0541 & 0.9893 & 0.0545 \\
$\beta_1$ & 0.25 & 0.2506 & 0.0033 & 0.2505 & 0.0033 \\
$\gamma_0$ & 0 & -0.0879 & 0.0926 & -0.0862 & 0.0931 \\
$\gamma_1$ & 5 & 4.8594 & 0.1861 & 4.8568 & 0.1836 \\
\end{tabular}
\end{table}

\subsection{Gaussian model with different scale parameters}
\label{s:gaussian}

In the last simulation study we have considered the case of grouped Gaussian
data so that each group has a different precision and the log-precision is
modeled on a mixed-effects model. In particular, we consider the model:

$$
\begin{array}{rcl}
Y_{ij} &\sim & N(\mu_{ij}, \tau_i);\ i=1,\ldots,p;\ j=1,\ldots,n_i\\
\mu_{ij} &= & \beta_0 + \beta_1 x_{ij}\\
\log(\tau_i) &=& \gamma_0 + \gamma_1 z_i+u_i\\
u_i &\sim& N(0, \tau_u)\\
\tau_u &\sim& Gamma(1, 0.00005)\\
\beta_0,\beta_1 &\sim& N(0, 0.001)\\
\gamma_0,\gamma_1 &\sim& N(0, 0.001)\\
\end{array}
$$
Here, $p$ represents the number of groups and $n_i$ the number of observations
in group $i$. The values of the parameters used in the simulations are $\beta_0 = 1$,
$\beta_1 = 0.25$, $\gamma_0 = 0$, $\gamma_1 = 5$ and $\tau_u = 1$. The total
number of observations is 2500, which corresponds to $p = 5$ groups and $n_i = 500,\
i=1,\ldots, p$.  Furthermore, values of covariate $x_{ij}$ have been simulated
from a uniform distribution between 0 and 1, while values of covariate $z_i$
have been obtained by sampling from a uniform distribution in the interval (-1,
1).

This model is a bit more complex because the log-precision depends on both
fixed and random effects. Hence, conditioning on $\bm\gamma$ alone will not
suffice to make this model a latent GMRF. It would be possible to condition on
$\bm\gamma$ and $\mathbf{u} = (u_i,\ldots, u_p)$ but then the dimension of the
parametric space may be difficult to handle by AMIS (in
particular, when the value of the number of groups $p$ is large). Furthermore,
estimating the random effects $u_i$ using importance sampling may be difficult,
and we prefer INLA to perform this task.

Instead, conditioning will be on $\bm\theta_c =
\bm\tau=(\tau_1,\ldots,\tau_p)$, which will split the main model into two
independent submodels with response variables $\mathbf{y}$ and $\log(\bm\tau)$,
as illustrated in Figure~\ref{fig:graphmodels} (bottom plot). These two models
can be fit independently and the resulting log-marginal likelihood will be the
sum of the corresponding values from the two models, which can be then used to
compute the weights.

Note that, in this particular case, nodes $\tau_1,\ldots,\tau_p$ are no
stochastic nodes as they are fully determined by $\bm\gamma$, $z_i$ and $u_i$.
For this reason, there is no prior for them. In order to ease the computations, and
without loss of generality, we set $\pi(\tau_i) = 1,\ i=1,\ldots,p$, which will not
have any effect on the computation of the marginal likelihood.

It is worth mentioning that, among the three different examples provided in the
simulation study, this one is an actual DHGLM as defined in
\cite{LeeNelder:2006} because it includes random effects when modeling
$\log(\tau_i)$.  In order to explore convergence of the AMIS algorithm we have
repeated the analysis using different sets of initial values for the parameters
of the importance distribution and number of samples (see below).  This will
allow us to explore how the adaptive procedure in the AMIS algorithm
behaves and to assess the resulting estimates precision.



Table~\ref{tab:gaussian_mcmc} summarizes the results of the estimation of the Gaussian model with the MCMC method.

\begin{table}[h!]
	\caption{Summary of the estimates of the Gaussian model with different scale parameters used in the simulation study, obtained by fitting the model with MCMC.}
	\label{tab:gaussian_mcmc}
	\centering
	\begin{tabular}{c|cccc}
		\multicolumn{4}{|c|}{MCMC} \\
		\hline
		Parameter & True value & Mean & St. dev. & 95 \% CI\\
		\hline
		$\beta_0$ & 1 & 0.9884 & 0.0170 & (0.9553, 1.0226) \\
		$\beta_1$ & 0.25 & 0.2864 & 0.0288 & (0.2284, 0.3421) \\
		$\gamma_0$ & 0 & -0.2926 & 0.3766 & (-1.0764, 0.4184) \\
		$\gamma_1$ & 5 & 3.8365 & 0.7832 & (2.2486, 5.3023) \\
		$\tau_u$ & 1 & 1.9128 & 1.2140 & (0.3247, 4.7935) \\
	\end{tabular}
\end{table}

Similarly, Table ~\ref{tab:gaussian_amis} shows the results of the estimation of the Gaussian model with the AMIS-INLA method, where different scenarios are considered. These scenarios are:
 \begin{enumerate}
     \item  Initial step of 5000 iterations, 10 new adaptive steps with 1000 simulations each, vague initial parameters for the sampling distribution (AMIS-INLA1).
     \item  Initial step of 5000 iterations, 10 new adaptive steps with 1000 simulations each, parameters informed from the data for the sampling distribution (AMIS-INLA2).
     \item  Initial step of 1000 iterations, 10 new adaptive steps with 1000 simulations each, vague initial parameters for the sampling distribution (AMIS-INLA3).
     \item  Initial step of 1000 iterations, 10 new adaptive steps with 1000 simulations each, parameters informed from the data for the sampling distribution (AMIS-INLA4).
     \item  Initial step of 5000 iterations, 10 new adaptive steps with 1000 simulations each, vague initial parameters and large variance for the sampling distribution (AMIS-INLA5).
     \item  Initial step of 5000 iterations, 10 new adaptive steps with 5000 simulations each, parameters informed from the data for the sampling distribution (AMIS-INLA6).
 \end{enumerate}

In all the scenarios described above, the sampling distribution is a
multivariate normal distribution for $(\log(\tau_1), \ldots,\log(\tau_p))$.
Vague initial parameters refers to using a mean of 0 and a variance matrix that
is diagonal with all entries equal to 5. Using a sampling distribution with
parameters informed from the data refers to computing the sample variance of each
group and computing the parameters of the sampling distribution from them. In
particular, the mean is the log of the vector of sample variances and the
variance matrix is diagonal with entries the variance of the log-sample
variances divided by their corresponding values of $n_i$.  If
the scenario indicated that a larger variance for the sampling distribution has
been used, these values are multiplied by 10.  In all cases these are initial
values of the parameters of the sampling distribution and they will be updated
at each adaptive step.

\begin{table}[h!]
	\caption{Summary of the estimates of the Gaussian model with different scale parameters used in the simulation study, obtained by fitting the model with the AMIS algorithm and INLA, for the six different scenarios considered}
	\label{tab:gaussian_amis}
	\centering
	\vspace{0.5cm}
	\begin{tabular}{c|cccc}
		\multicolumn{5}{|c|}{AMIS-INLA1} \\
		\hline
		Parameter & True value & Mean & St. dev. & 95 \% CI\\
		\hline
		$\beta_0$ & 1 & 0.9877 & 0.0174 & (0.9536,1.0216) \\
        $\beta_1$ & 0.25 & 0.2874 & 0.0299 & (0.2288,0.3458) \\
        $\gamma_0$& 0 & -0.3628 & 0.4062 & (-1.1867,0.4568) \\
        $\gamma_1$ & 5 & 3.5299 & 0.7874 & (1.9312,5.1173) \\
        $\tau_u$ & 1 & 2.1079 & 1.3224 & (0.3525,5.4083) \\
	\end{tabular}

    \vspace{1cm}
    \begin{tabular}{c|cccc}
    	\multicolumn{5}{|c|}{AMIS-INLA2} \\
    	\hline
    	Parameter & True value & Mean & St. dev. & 95 \% CI\\
    	\hline
    	$\beta_0$ & 1 & 0.9883 & 0.0163 & (0.9563,1.0201) \\
        $\beta_1$ & 0.25 & 0.2866 & 0.0280 & (0.2317,0.3414) \\
        $\gamma_0$ & 0 & -0.2620 & 0.4270 & (-1.1280,0.5994) \\
        $\gamma_0$ & 5 & 3.9200 & 0.8279 & (2.2390,5.5881) \\
        $\tau_u$ & 1 & 1.9268 & 1.2212 & (0.3193,4.9866) \\
    \end{tabular}

    \vspace{1cm}
    \begin{tabular}{c|cccc}
    	\multicolumn{5}{|c|}{AMIS-INLA3} \\
    	\hline
    	Parameter & True value & Mean & St. dev. & 95 \% CI\\
    	\hline
       $\beta_0$ & 1 & 0.9778 & 0.0244 & (0.9299,1.0257) \\
       $\beta_1$ & 0.25 & 0.3028 & 0.0420 & (0.2203,0.3850) \\
       $\gamma_0$ & 0 & -0.9351 & 0.8263 & (-2.6105,0.7322) \\
       $\gamma_1$ & 5 & 3.0919 & 1.5998 & (-0.1586,6.3142) \\
       $\tau_u$ & 1 & 0.5086 & 0.3189 & (0.0852,1.3045) \\
    \end{tabular}

    \vspace{1cm}
    \begin{tabular}{c|cccc}
    	\multicolumn{5}{|c|}{AMIS-INLA4} \\
    	\hline
    	Parameter & True value & Mean & St. dev. & 95 \% CI\\
    	\hline
    	$\beta_0$ & 1 & 0.9882 & 0.0163 & (0.9562,1.0202) \\
        $\beta_1$ & 0.25 & 0.2866 & 0.0280 & (0.2316,0.3415) \\
        $\gamma_0$ & 0 & -0.2625 & 0.4269 & (-1.1282,0.5987) \\
        $\gamma_1$ & 5 & 3.9173 & 0.8276 & (2.2369,5.5849) \\
        $\tau_u$ & 1 & 1.9282 & 1.2221 & (0.3195,4.9903) \\
    \end{tabular}

    \vspace{1cm}
    \begin{tabular}{c|cccc}
    	\multicolumn{5}{|c|}{AMIS-INLA5} \\
    	\hline
    	Parameter & True value & Mean & St. dev. & 95 \% CI\\
    	\hline
   	    $\beta_0$ & 1 & 0.9882 & 0.0163 & (0.9562,1.0201) \\
        $\beta_1$ & 0.25 & 0.2866 & 0.0280 & (0.2316,0.3415) \\
        $\gamma_0$ & 0 & -0.2630 & 0.4272 & (-1.1295,0.5987) \\
        $\gamma_1$ & 5 & 3.9173 & 0.8282 & (2.2356,5.5861) \\
        $\tau_u$ & 1 & 1.9254 & 1.2205 & (0.3190,4.9835) \\
    \end{tabular}

    \vspace{1cm}
    \begin{tabular}{c|cccc}
    	\multicolumn{5}{|c|}{AMIS-INLA6} \\
    	\hline
    	Parameter & True value & Mean & St. dev. & 95 \% CI\\
    	\hline
   	    $\beta_0$ & 1 & 0.9882 & 0.0163 & (0.9562,1.0201) \\
        $\beta_1$ & 0.25 & 0.2866 & 0.0280 & (0.2316,0.3415) \\
       $\gamma_0$ & 0 & -0.2625 & 0.4270 & (-1.1286,0.5989) \\
       $\gamma_1$ & 5 & 3.9170 & 0.8279 & (2.2360,5.5852) \\
       $\tau_u$ & 1 & 1.9273 & 1.2220 & (0.3193,4.9894) \\
    \end{tabular}

\end{table}

\begin{figure}[h!]
\caption{Estimates of the posterior marginals of the parameters obtained by fitting the Gaussian model with different scale parameters, using both the MCMC and AMIS-INLA methods, considering all scenarios for the AMIS-INLA algorithm setup.}
\includegraphics[width=\linewidth]{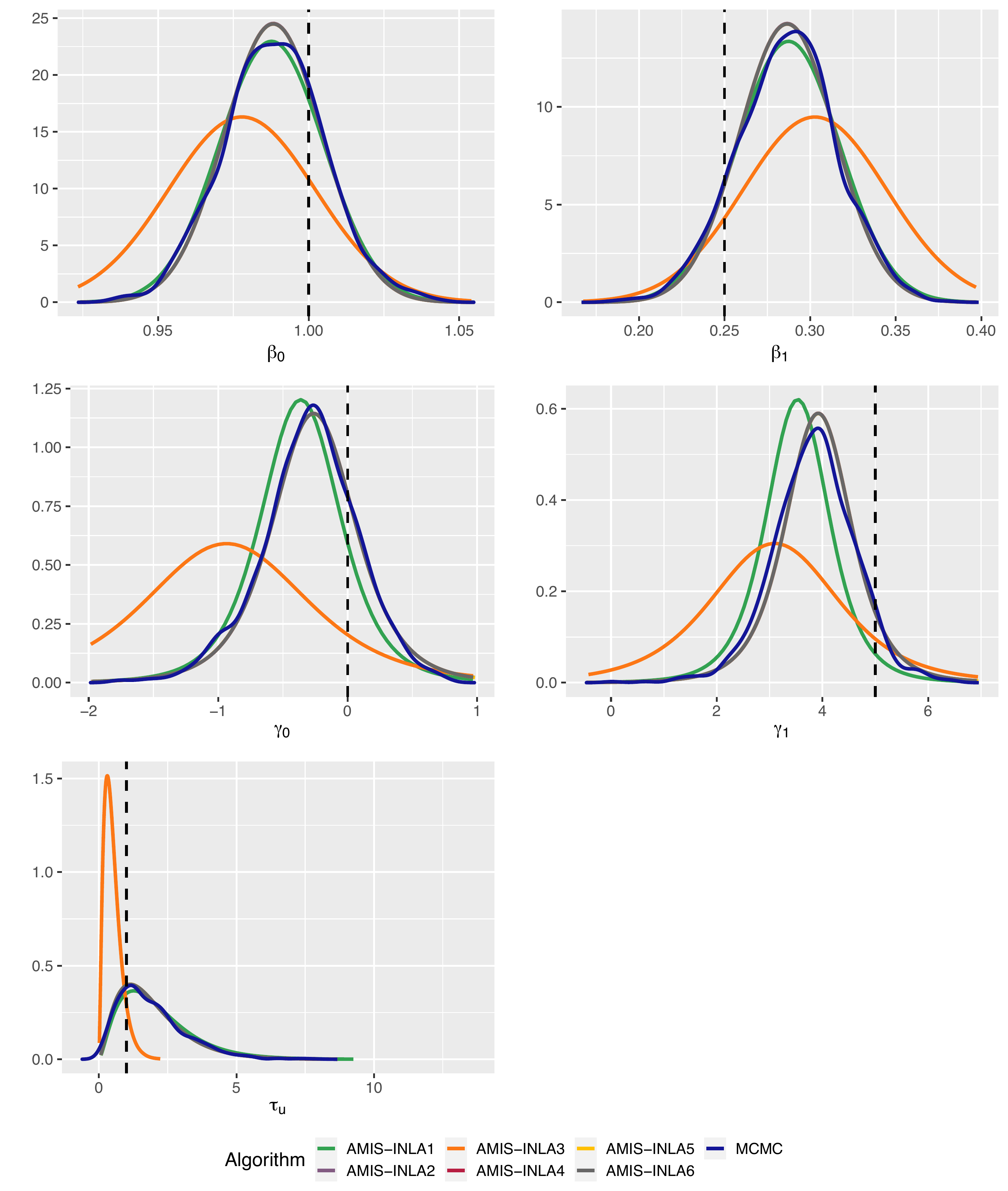}
\label{fig:gaussian_all}
\end{figure}

Furthermore, Figure~\ref{fig:gaussian_all} shows the estimates of the posterior
marginals of the parameters obtained with MCMC and the different settings of
the AMIS-INLA algorithm. Estimation is good for all model parameters for most
scenarios, with point estimates close to that of MCMC in most cases. However,
estimates of $\tau_u$ do not seem to be good as AMIS with INLA tends to
underestimate this parameter for scenario 3. The effective sample sizes of AMIS
with INLA range from 5.12 (scenario 3, based on 11000 simulations) to
10444.68 (scenario 2, 15000 total simulations) and 51536.74 (scenario 6, based
on a total of 55000 simulations). Hence, scenario 3 is likely to
produce poor estimates due to its low effective sample size.

It is worth noting that the estimation of the posterior marginal of $\tau_u$
has been conducted by first averaging the posterior marginal of $\log(\tau_u)$
(the internal scale of this parameter in INLA) and then transforming the
resulting marginal to obtain that of $\tau_u$. The reason is that INLA
estimates of the posterior marginal of $\tau_u$ were not reliable.

%
%

\subsection{Summary of results}
\label{s:summary}

The simulated studies conducted above illustrate the use of AMIS with INLA
to fit DHGLM. This approach will allow a flexible definition of the models using the
R-INLA package as well as efficient model fitting. Given that AMIS can be run
in parallel, DHGLM could be fit in a short time provided a computer with a
large number of CPUs is available (which is not uncommon these days).

Regarding the selection of the parameters in $\bm\theta_c$, we have provided
new guidelines not discussed in \cite{GomezRubioRue:2018} or \cite{Berildetal:2021}
by using the graphical representation of the models in
Figure~\ref{fig:graphmodels}. By inspecting the graphical model, it is
easier to find the parameters to condition on so that the resulting model
is a latent GMRF (see Poisson and negative binomial models). Furthermore,
for highly structured models, it is possible to split the model into more
than one submodel (that are latent GMRF) by conditioning on a small sample
of hyperparameters, as is the case of the Gaussian model with different scale parameters.

The parameters in $\bm\theta_c$ have been included in a red dotted box, which
has been labelled AMIS as this is the method used to estimate the
posterior distribution of these parameters. Similarly, the conditional latent
GMRF has been included in a blue dotted line, which has been labelled as
INLA because this is the method used to estimate the posterior marginals of
the parameters in this conditional model.

In a nutshell, the parameters in $\bm\theta_c$ should be taken so that
their dimension is as low as possible, preferable as part of coefficients of
fixed effects or precisions of random effects, and so that they split
the main model into one or more submodels that are easy to fit with INLA.
Choosing the random effects themselves as part of $\bm\theta_c$ should be avoided as it is
difficult to sample efficiently using AMIS and their dimension is likely to
increase with the size of the data.

\section{Examples}
\label{s:examples}

In this Section we illustrate model fitting of DHGLM with AMIS-INLA using
two real datasets.  Section~\ref{ex:colombia} describes a Poisson
model with random effects with a hierarchical structure on the precision and also a negative binomial model with a hierarchical structure on the size parameter
to analyze infant mortality in Colombia. Section~\ref{ex:sleep} fits a model
with subject-level random slopes and precisions to participants in a
sleep deprivation study.

\subsection{Infant mortality in Colombia}
\label{ex:colombia}

%

The infant mortality data in Colombia that we analyze here has been
studied in previous works \citep[see, for
example,][]{QuinteroSarmientoetal:2012,cepeda:2018,morales:2021}. The variables
available in this dataset are given for each of the $n=32$ departments or
regions of Colombia: the number of children under one year of age who
died in year 2005 (\textbf{ND}), the total number of births in the same
year (\textbf{NB}), an index that represents the percentage of people with
their basic needs not satisfactorily attended for year 2005 (\textbf{IBN})
and the observed mortality rates, computed as the number of children under one
year of age who died in 2005 per 1000 born alive (\textbf{Rates}).

It has been shown in previous works \citep[e.g.,][]{QuinteroSarmientoetal:2012}
that these data presents overdispersion when fitting a Poisson regression model
for the mortality rates, a phenomenon that arises when the real variance of the
data is larger than the one specified in the model. Additionally, there have
been findings of the evidence that there is spatial autocorrelation present in
the data \citep{cepeda:2018}. Therefore, these are issues that need to be taken
into account if we wish to specify regression models for this data.

The first model considered is the generalized spatial conditional normal
Poisson \citep{cepeda:2018}, which is able to accommodate overdispersion and to
explain spatial dependence. This model assumes that the variable representing
the number of deaths in each region ($\rmn{ND}_{i}$), conditioned on the set of
values it takes in the neighboring regions without including region $i$ itself
($\rmn{ND}_{\sim i}$) and on a set of normally distributed random effects
$u_{i} \sim N(0,\tau_{i})$ follows a Poisson distribution, that is
$(\rmn{ND}_i| \rmn{ND}_{\sim i}, u_i) \sim \rmn{Poi}(\mu_i)$ for $i=1, \dots ,
n$.

This model allows the dispersion parameter to vary according to explanatory variables or
any other terms by specifying a regression model for the variance of the random effect.
It is also able to explain the spatial association which may be present in the data by
including the spatial lag of the rates in the regression model for the mean or in the model
for the dispersion as well \citep[see][]{cepeda:2018,morales:2021}.

The connection with DHGLM appears here because we can model the
log-precisions using a linear predictor on \textbf{IBN} so that $\log(\tau_i) =
\gamma_{0} + \gamma_{1}\rmn{IBN}_{i}, i=1,\ldots, n$. It is worth mentioning
that, in this particular case, the precisions are univocally determined by the linear
predictor.

Following the example from \cite{morales:2021}, we have specified the following model:

$$
\begin{array}{rcl}
(\rmn{ND}_i \mid \rmn{ND}_{\sim i}, u_i) & \sim & \rmn{Poi}(\mu_i) \\
 \log(\mu_{i})  & = & \log(\rmn{NB}_i) + \beta + \rho \mathbf{W}_{i}\mathbf{Rates} + u_{i} \\
      u_{i} &\sim & N(0,\tau_{i}) \\
 \log(\tau_{i}) & = & \gamma_{0} + \gamma_{1}\rmn{IBN}_{i} \\
 \beta,\rho &\sim& N(0, 0.001)\\
 \gamma_0,\gamma_1 &\sim& N(0, 0.001),\\
\end{array}
$$
where $\mathbf{W}_{i}$ is the $i$-th row of a row-standardized spatial
neighborhood matrix $\mathbf{W}$. Adjacency here is defined so that
two regions are neighbours if they share at least one point of their boundaries.
Therefore, $\mathbf{W}_{i}\mathbf{Rates}$ is the
spatial lag of the observed mortality rates, which in this case represents
the average of $\mathbf{Rates}$ at the neighbours.

In the implementation of AMIS with INLA we have taken $\bm\theta_c = \bm\gamma
= (\gamma_0, \gamma_1)$. The sampling distribution is a bivariate Gaussian
with vector mean $(0, 0)$ and the variance matrix is a diagonal matrix with
entries equal to 5. In this case, 5000 simulations were initially run, followed by
10 adaptive steps with 1000 simulations each.

\begin{figure}[h!]
	\caption{Posterior marginals of the estimated parameters obtained by fitting the generalized spatial conditional normal Poisson model to the infant mortality data in Colombia, using both the MCMC and AMIS-INLA methods.}
	\includegraphics[width=\linewidth]{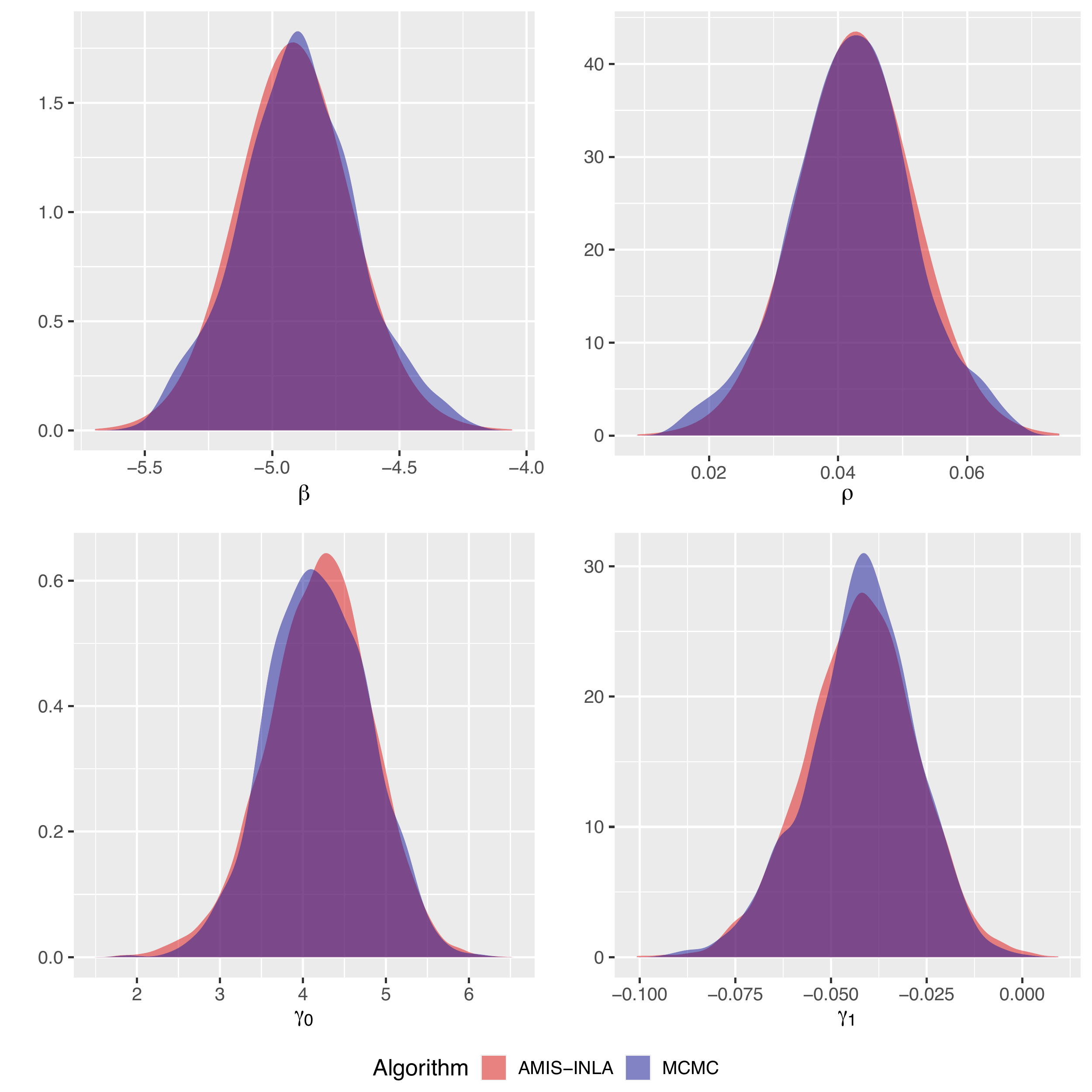}
	\label{fig:dens_colpoi}
\end{figure}

Results of the estimation of this model are shown in Table~\ref{tab:col_poi}
and Figure~\ref{fig:dens_colpoi}. As can be seen, AMIS-INLA and MCMC
produce close results. The effective sample size of AMIS with INLA is 9263.002.

\begin{table}[h!]
	\caption{Summary of the estimates of the generalized spatial conditional normal Poisson model with random effects and varying dispersion fitted to the infant mortality data in Colombia.}
	\label{tab:col_poi}
	\centering
	\begin{tabular}{c|cc|cc}
		& \multicolumn{2}{|c|}{AMIS} & \multicolumn{2}{|c}{MCMC}\\
		\hline
		Parameter & Mean & St. dev. & Mean & St. dev.\\
		\hline
		$\beta$ & -4.9124 & 0.2306 & -4.8987 & 0.2310 \\
        $\rho$ & 0.0427 & 0.0094 & 0.0421 & 0.0095 \\
        $\gamma_0$ & 4.1951 & 0.6392 & 4.1893 & 0.6033 \\
        $\gamma_1$ & -0.0423 & 0.0148 & -0.0421 & 0.0140 \\
	\end{tabular}
\end{table}

The negative binomial model could be another option to consider in order to fit the infant mortality data described here. Therefore, we have specified the generalized spatial conditional negative binomial model \citep{cepeda:2018}, where it is assumed that $(\rmn{ND}_i| \rmn{ND}_{\sim i}) \sim \rmn{NB}(\mu_i, \rmn{k}_i)$, with $\mu_{i}$ being the conditional mean and $\rmn{k}_{i}$ the size parameter of a negative binomial distribution. For this model, we can specify regression structures both for the mean and dispersion parameters, which can include the spatial lag of the rates and explanatory variables as well.

In particular, we have fitted the following model:

$$
\begin{array}{rcl}
(\rmn{ND}_i \mid \rmn{ND}_{\sim i}) & \sim & \rmn{NB}(\mu_i, \rmn{k}_i) \\
\log(\mu_{i})  & = & \log(\rmn{NB}_i) + \beta + \rho \mathbf{W}_{i}\mathbf{Rates}\\
\log(\rmn{k}_{i}) & = & \gamma_{0} + \gamma_{1}\rmn{IBN}_{i} \\
\beta,\rho &\sim& N(0, 0.001)\\
\gamma_0,\gamma_1 &\sim& N(0, 0.001)\\
\end{array}
$$

In order to fit this model with AMIS with INLA we have also taken $\bm\theta_c =
\bm\gamma = (\gamma_0, \gamma_1)$. Conditional on $\bm\theta_c$, the resulting
model is a negative binomial with different known sizes, which is easy to fit
with INLA. Sampling has been done as with the Poisson distribution.

\begin{figure}[h!]
	\caption{Posterior marginals of the estimated parameters obtained by fitting the generalized spatial conditional negative binomial model to the infant mortality data in Colombia, using both the MCMC and AMIS-INLA methods.}
	\includegraphics[width=\linewidth]{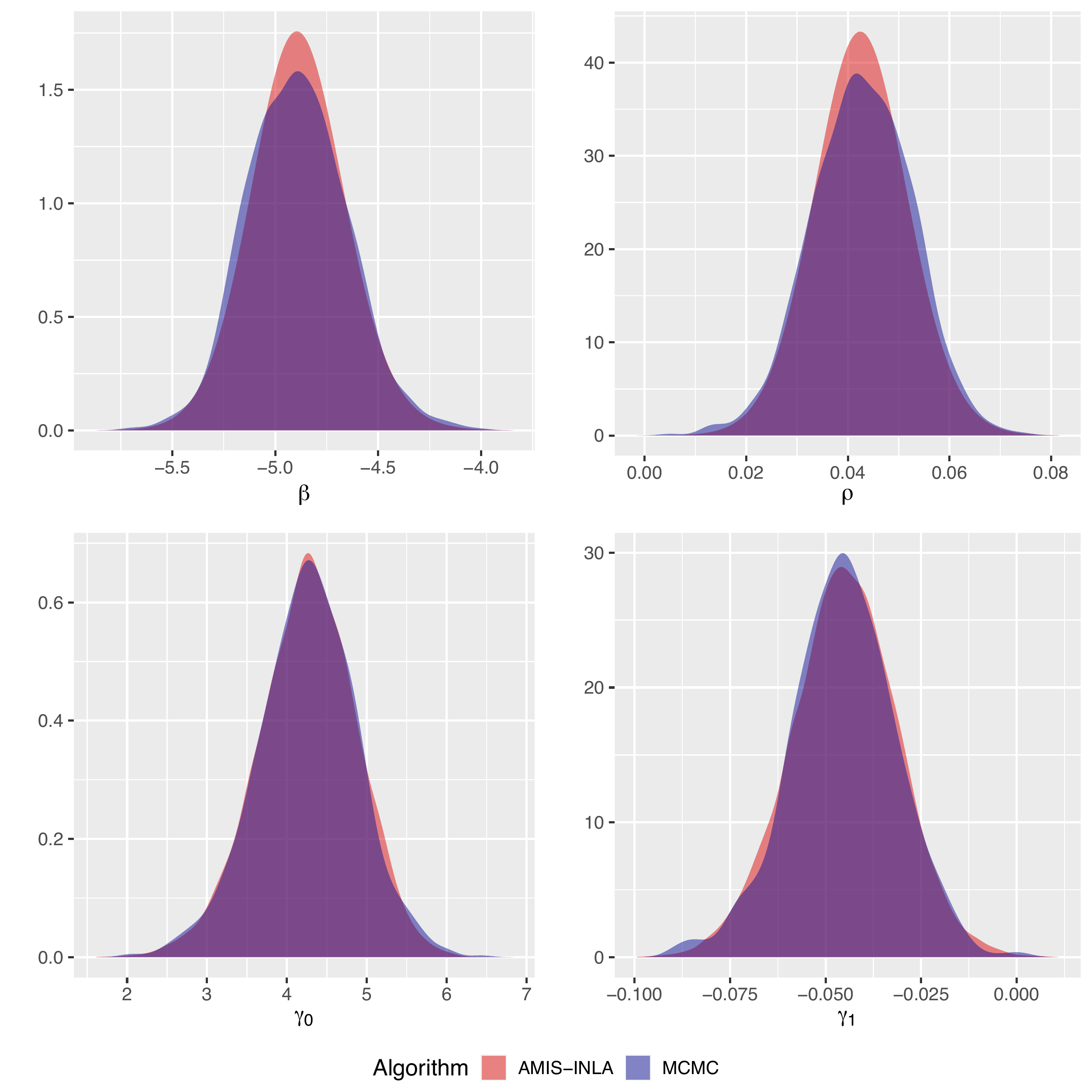}
	\label{fig:dens_colnb}
\end{figure}

Table~\ref{tab:col_nb} and Figure~\ref{fig:dens_colnb} display the 
results of the estimation of this model, which show that AMIS with INLA 
provides very similar results to MCMC. The effective sample
size of AMIS with INLA is 9717.207 now.

\begin{table}[h!]
	\caption{Summary of the estimates of the generalized spatial conditional negative binomial model with varying dispersion fitted to the infant mortality data in Colombia.}
	\label{tab:col_nb}
	\centering
	\begin{tabular}{c|cc|cc}
		& \multicolumn{2}{|c|}{AMIS-INLA} & \multicolumn{2}{|c}{MCMC}\\
		\hline
		Parameter & Mean & St. dev. & Mean & St. dev.\\
		\hline
		$\beta$ & -4.8871 & 0.2341 & -4.8933 & 0.2427 \\
        $\rho$ & 0.0425 & 0.0094 & 0.0423 & 0.0099 \\
        $\gamma_0$ & 4.2547 & 0.6235 & 4.2553 & 0.6191 \\
        $\gamma_1$ & -0.0452 & 0.0142 & -0.0454 & 0.0139 \\
	\end{tabular}
\end{table}

\subsection{Sleep deprivation study}
\label{ex:sleep}

\cite{Belenkyetal:2003} conducted an experiment to measure the effect of sleep
deprivation on reaction time on a number of subjects.  A subset of this dataset
is included in the \texttt{R} package \texttt{lme4} \citep{lme4} and it includes
observations for the most sleep-deprived group for the first 10 days of the
study. This dataset has been analyzed by different authors \citep[see, for
example,][]{GomezRubio:2020} using linear mixed-effects with random slopes as
the number of days under sleep deprivation seems to have a different effect on
the different subjects.

\begin{figure}[h!]
\caption{Effect of number of days under sleep deprivation on different subjects (based on code from the \texttt{lme4} package).}
\includegraphics[width=\linewidth]{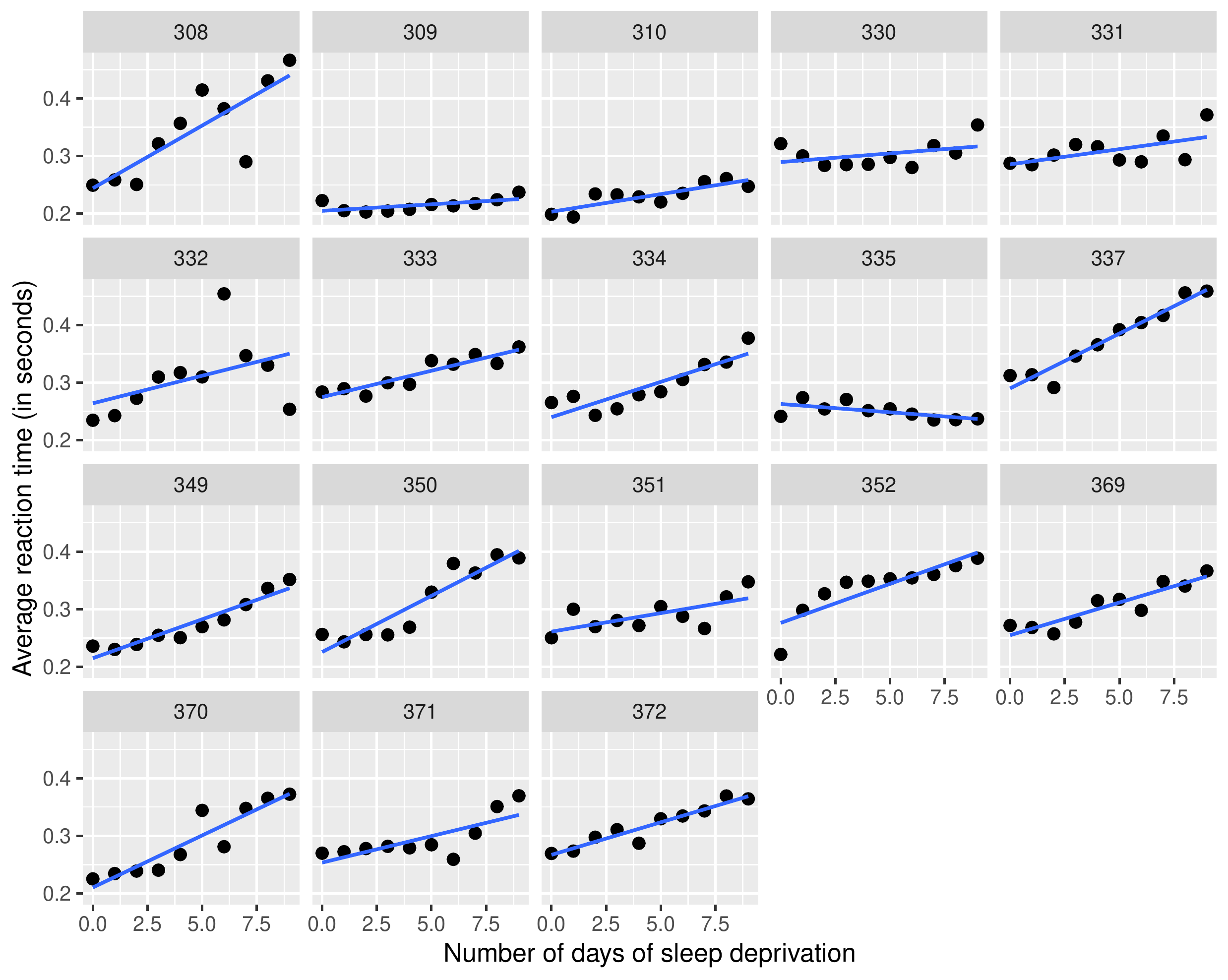}
\label{fig:sleepdata}
\end{figure}

Subject-specific reaction times accompanied by their respective linear
regression lines can be seen in Figure~\ref{fig:sleepdata}. This figure also
illustrates the fact that variability of the reaction times among subjects is
not uniform, with some subjects having a broader range of values than others.
For this reason, we have fitted a model with random slopes per subject in which
the within-in subject precision of the measurements is different using a DHGLM.

In particular, we have fitted the following model:

\begin{equation}
\begin{array}{rcl}
Y_{ij} &\sim & N(\mu_{ij}, \tau_i);\ i=1,\ldots,p;\ j=1,\ldots,n_i\\
\mu_{ij} &= & \beta_0 + \beta_i \rmn{day}_{ij}\\
\log(\tau_i) &=& \gamma + u_i\\
\beta_i &\sim& N(0, \tau_{\beta})\\
u_i &\sim& N(0, \tau_u)\\
\tau_{\beta} &\sim& Gamma(1, 0.00005)\\
\tau_u &\sim& Gamma(1, 0.00005)\\
\beta_0 &\sim& N(0, 0.001)\\
\gamma &\sim& N(0, 0.001)\\
\end{array}
\label{eq:sleepstudy_rcoef}
\end{equation}

Here, $p = 18$ is the number of subjects and $n_i=10,\ i=1,\ldots, p$ given that
all subjects have the same number of measurements in the dataset. Covariate
$\rmn{day}_{ij}$ is the number of the days since the beginning of the sleep deprivation
experiment. Note that $\beta_i, i=1,\ldots,p$ refers to random coefficients
to allow for different per-subject slopes. It should be emphasized that this model is similar
to the one in Section~\ref{s:gaussian} and that it will be fitted in a similar
way, i.e., by sampling from $(\log(\tau_1), \ldots, \log(\tau_p))$.
Note that the dimension of the parametric space is 18, which may be large
for algorithms such as IS and AMIS.

In order to select the parameters of the importance distribution we have
proposed different approaches. Initially, we assumed a multivariate normal distribution
with zero mean and a diagonal precision matrix with entries equal to 5 along
the diagonal. This provided a vague starting sampling distribution for the
log-precisions that after a few adaptation steps may get close to the actual
posterior distribution. Unfortunately, this provided very poor estimates
and the results were discarded.

We noticed that importance sampling may not be efficient if the mean of the
importance distribution is far from the posterior modes and also when its
variance is too large. For this reason, we propose to use the data to obtain
some rough estimates of the posterior mean and precisions based on $S^2_i$, the
sample variance computed using measurements from subject $i$. Then, the mean of
the importance distribution is $(\log(1 / S^2_1), \ldots, \log(1 / S^2_p))$ and
the variance is diagonal with entries $0.05 \cdot (\log(1 / S^2_1), \ldots,
\log(1 / S^2_p))$.  In principle, this should provide a starting sampling
distribution which is close to the posterior modes and with a variance in the
scale of the posterior variances that allows for short jumps during the
adaptive steps.

However, we noticed that we could obtain better initial parameters by
performing permutations of the values of $(\log(1 / S^2_1), \ldots, \log(1 /
S^2_p))$, fitting the conditional model and checking the values of the
conditional marginal likelihood, so that the permutation with the highest value
is used to set the parameters of the initial sampling distribution.  This
simple prior step produced means of the sampling distribution that were very
closed to the posterior mode of $(\log(\tau_1), \ldots, \log(\tau_p))$. In
particular, 500 random permutations were tested prior to running AMIS with
INLA.

For all the models fitted in this example, AMIS with INLA has been run using an
initial adaptive step based on 1000 simulations followed by 20 adaptive steps
with 1000 simulations each. MCMC is based on 10000 burn-in simulations followed
by 100000 simulations, of which only 1 in 100 has been kept, so that inference
is based on 1000 samples.

%

\begin{figure}[h!]
	\caption{Posterior marginals of the estimated parameters obtained by fitting the Gaussian model with random slopes for each subject to the sleep study data, using both the MCMC and AMIS-INLA methods.}
	\includegraphics[width=\linewidth]{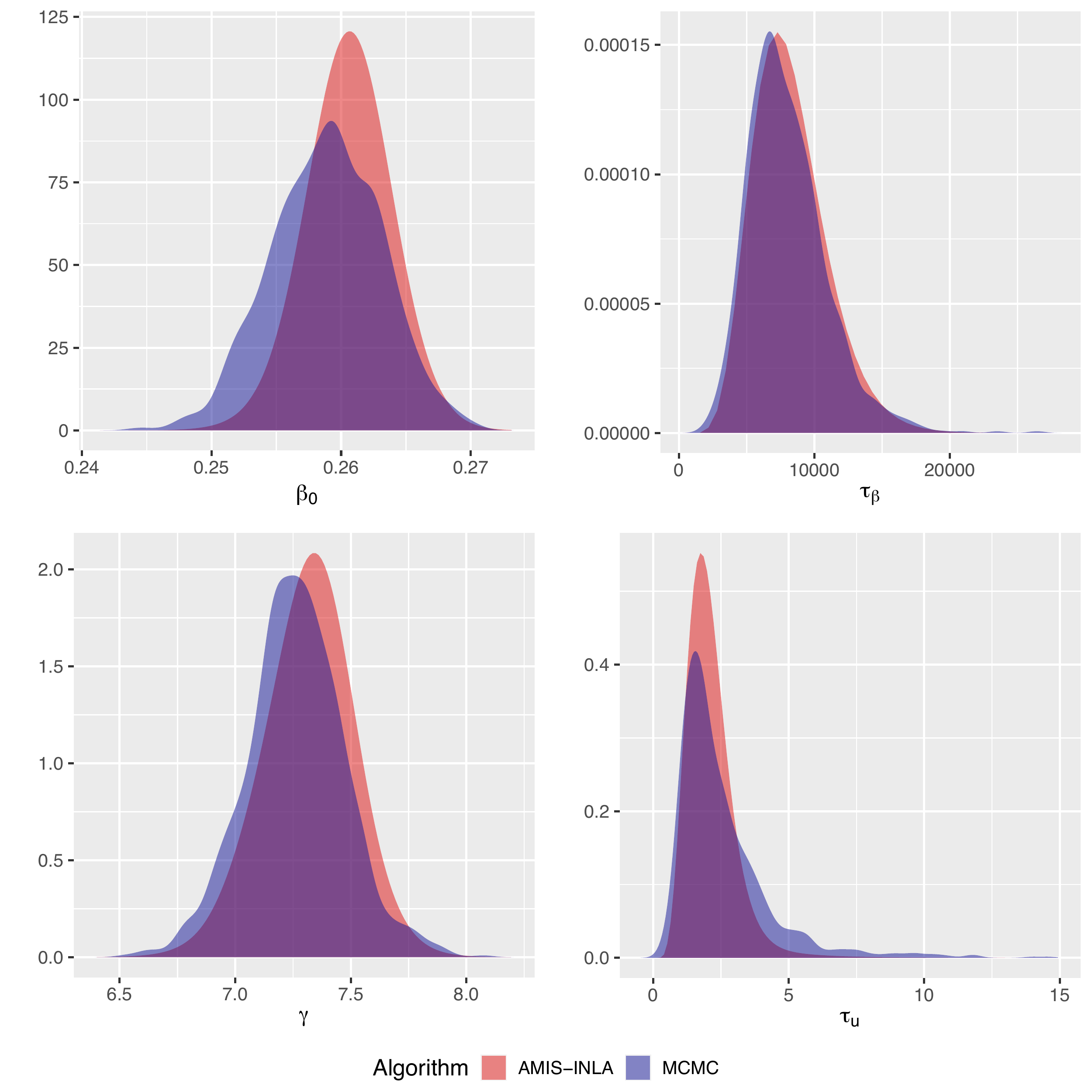}
	\label{fig:dens_sleepstudy_rcoef}
\end{figure}

Results of the estimation of this model are provided in Table~\ref{tab:sleepstudy_rcoef}
and the densities of the posterior estimations for the parameters are shown
in Figure~\ref{fig:dens_sleepstudy_rcoef}. The effective sample size
of AMIS with INLA in this case is 2.015619, which is small but seems to provide
good estimates of the marginals of the model parameters. It is worth mentioning
that we have computed the effective sample size after each adaptive step and that
it reached the value 81.14083 after 12 adaptation steps. AMIS with INLA could
be stopped after a certain effective sample size has been achieved. It is worth
noting that the estimates of $\log(\tau_i)$ did not change considerably in the last adaptive steps.

\begin{table}[h!]
	\caption{Summary of the estimates of the Gaussian model with random slopes for each subject fitted to the sleep study data.}
	\label{tab:sleepstudy_rcoef}
	\centering
	\begin{tabular}{c|cc|cc}
		& \multicolumn{2}{|c|}{AMIS} & \multicolumn{2}{|c}{MCMC}\\
		\hline
		Parameter & Mean & St. dev. & Mean & St. dev.\\
		\hline
        $\beta_0$ & 0.2606 & 0.0034 & 0.2589 & 0.0042 \\
        $\tau_{\beta}$ & 8240.2229 & 2742.75 & 8002.245 & 2898.704 \\
        $\gamma$ & 7.3170 & 0.2003 & 7.2612 & 0.2115 \\
        $\tau_{u}$ & 2.1222 & 0.9348 & 2.6565 & 2.2836 \\
	\end{tabular}
\end{table}

Furthermore, we have also considered a model with fixed effects:

\begin{equation}
\begin{array}{rcl}
Y_{ij} &\sim & N(\mu_{ij}, \tau_i);i=1,\ldots,p;\ j=1,\ldots,n_i\\
\mu_{ij} &= & \beta_0 + \beta_1 \rmn{day}_{ij}\\
\log(\tau_i) &=& \gamma + u_i\\
u_i &\sim& N(0, \tau_u)\\
\tau_u &\sim& Gamma(1, 0.00005)\\
\beta_0,\beta_1 &\sim& N(0, 0.001)\\
\gamma &\sim& N(0, 0.001)\\
\end{array}
\label{eq:sleepstudy}
\end{equation}

\begin{figure}[h!]
	\caption{Posterior marginals of the estimated parameters obtained by fitting the Gaussian model with fixed effects fitted to the sleep study data, using both the MCMC and AMIS-INLA methods.}
	\includegraphics[width=\linewidth]{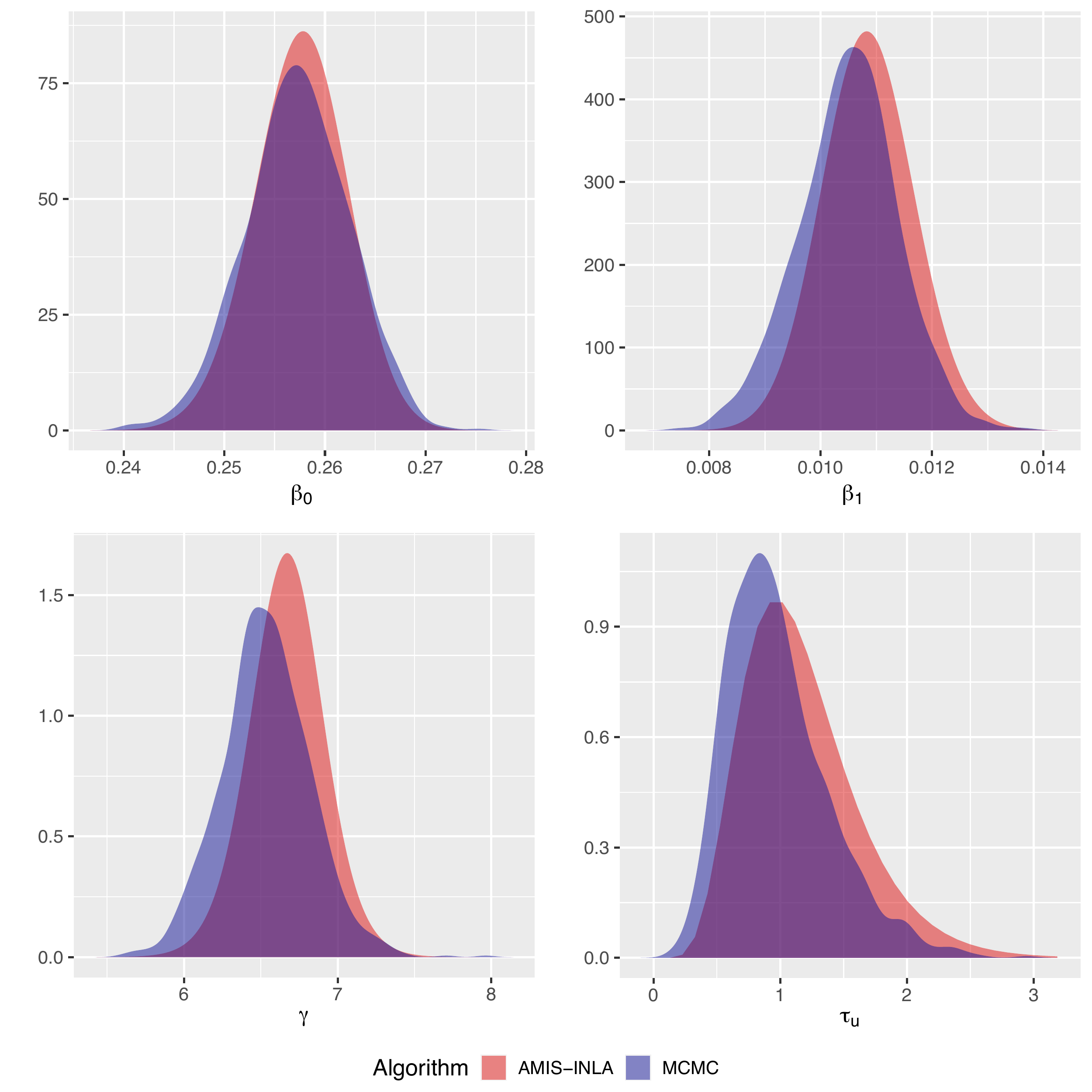}
	\label{fig:dens_sleepstudy}
\end{figure}

Results of the estimation of this model are provided in
Table~\ref{tab:sleepstudy} and the densities of the posterior estimations for
the parameters are shown in Figure~\ref{fig:dens_sleepstudy}.  Both models show
that, in general, AMIS with INLA and MCMC produce similar estimates.  In this
case, AMIS with INLA results an effective sample size of 71.906.

\begin{table}[h!] \caption{Summary of the estimates of the Gaussian model
fitted to the sleep study data.}
	\label{tab:sleepstudy}
	\centering
	\begin{tabular}{c|cc|cc}
		& \multicolumn{2}{|c|}{AMIS} & \multicolumn{2}{|c}{MCMC}\\
		\hline
		Parameter & Mean & St. dev. & Mean & St. dev.\\
		\hline
        $\beta_0$ & 0.2576 & 0.0047 & 0.2572 & 0.0051 \\
        $\beta_1$ & 0.0109 & 0.0008 & 0.0105 & 0.0009 \\
         $\gamma$ & 6.6680 & 0.2496 & 6.5348 & 0.2889 \\
         $\tau_u$ & 1.1676 & 0.4536 & 0.9794 & 0.4032 \\
	\end{tabular}
\end{table}

\section{Discussion}
\label{s:discuss}

Double hierarchical models present a particular structure that
models both the mean and scale parameter of different hierarchical models
with likelihood within the exponential family. Hence, inference on these models can be difficult
due to the different levels and effects in the model hierarchy.
We have illustrated how the integrated nested Laplace approximation can
be used to fit these models by using importance sampling
and adaptive multiple importance sampling.

In practice, this allows INLA to integrate most of the latent effects and
hyperparameters out so that a small subset of them is estimated using
importance sampling. Given that IS can be easily parallelized, this provides an
approach that is computationally competitive and computing times can be close
to the ones provided by INLA.

We have illustrated model fitting of DHGLM by conducting three different
simulation studies and the analysis of two real datasets. In all cases,
conducting an adaptive multiple importance sampling provided good estimates of
the model effects and hyperparameters that were similar to those obtained with
Markov chain Monte Carlo methods.

Although we have discussed examples with Gaussian, Poisson and negative binomial
data, the approach presented here can be applied to any of the distributions in
the exponential family and, more generally, to other likelihood distributions
that can be used together with the R-INLA software. Any model that
can be expressed as a latent GMRF by conditioning on a (small) subset
of latent effects or hyperparameters is susceptible to be fitted with
IS/AMIS with INLA.

Finally, the R code used to develop the simulation study and the examples
is available from \url{https://github.com/becarioprecario/DHGLM-INLA}. The data for the infant mortality in Colombia in Section~\ref{s:examples} have been replaced by a simulated dataset due to confidentiality constrains.


\backmatter


\section*{Acknowledgements}

V. N\'u\~{n}ez-Ant\'on and M. Morales-Otero's research has been funded by
Ministerio de Ciencia e Innovaci\'on (MCIN, Spain), Agencia Estatal de
Investigaci\'on (AEI/10.13039/501100011033/) and Fondo Europeo de Desarrollo
Regional (FEDER) ``Una manera de hacer Europa'' under the I+D+i research grant
PID2020-112951GB-I00 and by Ministerio de Econom\'{\i}a y Competitividad
(Spain), Agencia Estatal de Investigaci\'on (AEI), and the European Regional
Development Fund (ERDF), under research grant MTM2016-74931-P (AEI/ERDF, EU).
In addition, V. N\'u\~{n}ez-Ant\'on's research has also been funded by the
Department of Education of the Basque Government (UPV/EHU Econometrics Research
Group) under research grant IT-1359-19.

V. G\'omez-Rubio has been supported by grant SBPLY/17/180501/000491, funded by
Consejer\'ia de Educaci\'on, Cultura y Deportes (JCCM, Spain) and FEDER,
grant PID2019-106341GB-I00, funded by Ministerio de Ciencia e Innovaci\'on
(Spain) and a group support grant from Universidad de Castilla-La Mancha (Spain).

The code for the AMIS with INLA algorithm has been obtained from an early
version of the code described in \cite{Berildetal:2021} available from
\url{https://github.com/berild/inla-mc}.


%
\bibliographystyle{biom}
\bibliography{biblio.bib}


%
%
%
%
%

\label{lastpage}

\end{document}